\documentclass[aps,prd,twocolumn,showpacs,superscriptaddress,nofootinbib,preprintnumbers]{revtex4-2}
\usepackage{mathrsfs}
\usepackage{style}
\usepackage{diagbox}
\usepackage{booktabs}
\usepackage{multirow}
\usepackage{siunitx}
\usepackage{balance}

\newcommand{\bea}{\begin{equation}\begin{aligned}}
\newcommand{\eea}{\end{aligned}\end{equation}}

\begin{document}
\vspace{-1cm}
\hfill{\small \tt ~~~~~~~~~~~~~ FERMILAB-PUB-26-0113-T}
\hfill
\vspace{0.5cm}

\title{\textbf{New Thermal-Relic Targets for sub-GeV Dark Matter Direct Detection}}

\author{Xu Han} 
\email{xuhan7@uchicago.edu}
\affiliation{Department of Physics, University of Chicago, Chicago, IL 60637}

\author{Gordan Krnjaic} 
\email{krnjaicg@uchicago.edu}
\affiliation{Fermi National Accelerator Laboratory, Batavia, Illinois 60510}
\affiliation{Kavli Institute for Cosmological Physics, University of Chicago, Chicago, IL 60637}
\affiliation{Department of Astronomy and Astrophysics, University of Chicago, Chicago, IL 60637}

\date{\today}

\begin{abstract}
Dark matter direct detection experiments involving electron recoils are beginning to test highly-predictive, thermal-relic milestones for sub-GeV dark matter models. Due to the Lee-Weinberg bound, thermal dark matter candidates in this mass range necessarily require comparably-light mediator particles to achieve a suitably large annihilation cross section. Here we present new thermal-relic milestones for sub-GeV dark matter candidates that couple to vector mediators. In these models, the mediators are massive gauge bosons of anomaly-free abelian extensions to the Standard Model, including the dark photon, gauged $L_i - L_j, B-L$, and $B-3L_i$ models, where $B$ is the baryon number, $L$ is the lepton number, and $i,j$ index the lepton families. Since the same interactions that govern cosmological production also govern electron scattering, the targets we present are firmly predictive and allow for these models to be robustly discovered or falsified. Furthermore, since the mediators we study exhaust the minimal anomaly-free U(1) extensions to the Standard Model, our results offer a complete list of predictive milestones for sub-GeV dark matter coupled to vector mediators.
\end{abstract}

\vspace{-1cm}
\maketitle

\section{Introduction}

Thermal freeze out is a compelling and highly-predictive  mechanism for producing the dark matter (DM) abundance. In this framework, DM is initially in  equilibrium with the Standard Model (SM) plasma in the early universe~\cite{Cirelli:2024ssz}. As the universe cools, thermal contact is eventually lost and the observed relic density is achieved if the annihilation cross section satisfies $\langle \sigma v\rangle \sim 10^{-26}\,\text{cm}^3\text{s}^{-1}$~\cite{Steigman_2012}. Notably, this mechanism is insensitive to unknown high-energy physics and often depends only on model parameters that also govern laboratory observables.  

Freeze out can be realized across  the $\sim$ MeV-100 TeV DM mass range, where upper and lower bounds are set by unitarity~\cite{Griest:1989wd} and cosmology~\cite{Steigman_2012}, respectively. Light MeV-GeV thermal DM candidates have long been studied as viable alternatives to weak-scale dark matter~\cite{Boehm:2003hm}.
In the sub-GeV regime, the Lee-Weinberg bound requires comparably-light new ``mediators" to avoid overproducing DM~\cite{Lee:1977ua}. By far, the most commonly studied mediator is the dark photon whose SM interaction arises from kinetic mixing with the photon~\cite{Fabbrichesi_2021}, though several other possibilities are also viable~\cite{Bauer:2018onh, Barman:2024lxy, Barman:2025jbo, Bernal:2025szh, KA:2023dyz}.

Over the past decade, the direct-detection community has made great progress towards probing sub-GeV DM using novel target materials~\cite{Essig:2011nj,Essig:2012yx,Essig:2015cda,Lin:2019uvt,Mitridate:2022tnv}. Unlike traditional nuclear-recoil searches, which lose sensitivity to DM masses below the GeV scale, these new strategies typically utilize much lighter target particles (e.g., electrons), which are sensitive to smaller energy depositions and can therefore probe keV-GeV DM candidates. Collectively, these new experiments, including Damic-M~\cite{Castell_Mor_2020}, SENSEI~\cite{PhysRevLett.122.161801,SENSEI:2023gie}, and CRESST-III~\cite{cresstcollaboration2024observationsinglephotonscresst} among others are now beginning to test the thermal freeze out mechanism. 

Recently, the Damic-M collaboration has ruled out the most accessible thermal-relic model: complex scalar DM $\chi$, which directly annihilates to SM particles through a kinetically mixed dark photon\footnote{Technically it is possible to revive this model in a narrow range of parameter space if the annihilation is resonant with $ m_{A^\prime} \approx 2m_\chi$~\cite{Izaguirre:2015yja,Feng:2017drg}. Also the exclusion requires $m_{A^\prime} > m_
\chi$ to ensure that the annihilation is $s$-channel and depends on the kinetic mixing.}~\cite{DAMIC-M:2025luv,krnjaic2025testingthermalrelicdarkmatter}. Although there are other viable models with dark photon mediators (e.g., with Majorana or pseudo-Dirac fermion DM candidates)~\cite{Berlin:2018bsc, PhysRevD.111.056003}, their direct-detection cross sections are either velocity- or loop-suppressed, so future accelerator searches will be necessary to test these scenarios~\cite{Krnjaic:2022ozp}. 

In this paper we go beyond the dark-photon mediator and identify new sub-GeV thermal-relic targets that are accessible with current and future direct detection experiments. Our focus is on mediators that arise from anomaly-free gauged $U(1)$ extensions to the SM: $L_i-L_j, B-L,$ and $B-3L_i$ where $B$ is the baryon number, $L$ is the lepton number, and $i,j$ are family indices. 

We couple these mediators to complex-scalar or Dirac fermion DM candidates and identify predictive parameter space in which the cosmological 
abundance is in one-to-one correspondence with the scattering cross section, thus enabling discovery or falsification with sufficient experimental sensitivity. We find that, models with tree-level electron couplings are severely constrained by current limits, but that electrophobic mediators leave large regions of predictive thermal parameter space remain viable and within reach of future experimental searches. 

This paper is organized as follows: In Sec.~\ref{sec:models}, we introduce the theoretical models and mediators. In Sec.~\ref{sec:relic}, we detail the thermal freeze-out mechanism. We discuss the relevant experimental and observational constraints in Sec.~\ref{sec:constraints}, before presenting our numerical results in Sec.~\ref{sec:results}. Finally, in Sec.~\ref{sec:discussion}, we offer our concluding remarks.

\section{Model Overview}
\label{sec:models}

In this section we describe the models considered in this paper. 
For each case, the most general interaction Lagrangian can be written as 
\begin{eqnarray}
\label{eq:Lint}
    {\cal L}_{\rm int} \supset - Z^\prime_\mu \left(g_{Z'} J_{Z'}^\mu+ \varepsilon  e J^\mu_\text{EM} +
    g_D J^\mu_D \right),
\end{eqnarray}
where $Z^\prime$ is the gauge boson of a new abelian group, $J^\mu_{Z^\prime}$ is a current of SM particles charged under this group with a coupling $g_{Z^\prime}$, $\varepsilon$ is a kinetic mixing parameter, $e$ is the electric charge, and $J_{\rm EM}^\mu$ is the SM electromagnetic current. With a coupling $g_D$, here $J_D^\mu$ is the DM current and we consider two representative cases for each mediator 
\begin{eqnarray}
    J^\mu_D = 
    \begin{cases}
        i\chi^*\partial_\mu\chi + \text{h.c.}, & \text{Complex Scalar,} \\
        \bar\chi\gamma^\mu \chi, & \text{Dirac Fermion.}\\
    \end{cases}
\end{eqnarray}
While other DM candidates are also possible  (e.g., Majorana or pseudo-Dirac fermions),  we focus on the complex scalar and Dirac fermion cases since their scattering cross sections are unsuppressed in the non-relativistic limit, and therefore most promising for direct detection searches.\footnote{For Majorana fermions, the cross section is velocity-suppressed, and for pseudo-Dirac fermions with a sufficient mass splitting, the elastic cross section arises at loop level and is sharply suppressed~\cite{Berlin:2018bsc}.}

For $m_{Z^\prime} > 2m_\chi$, the mediator predominantly decays to DM pairs. In the complex scalar model, the width for this channel is 
\begin{eqnarray}
\label{eq:WidthScalar}
\Gamma_{Z'}=\frac{\alpha_D m_{Z'}}{12}\left(1-\frac{\! 4m_\chi^2}{\,\,\,
m_{Z'}^2}\right)^{3/2},
\end{eqnarray}
where $\alpha_{D} = g_{D}^2/4\pi.$ 
For Dirac fermions, the corresponding width is 
\begin{equation}
\label{eq:WidthDirac}
\Gamma_{Z'}=\frac{\alpha_D m_{Z'}}{3}\left(1+\frac{\! 2m_\chi^2}{\, \, \, m_{Z'}^2}\right)\sqrt{1-\frac{4m_\chi^2}{m_{Z'}^2}}.
\end{equation}
For each mediator, there are also SM decay channels, but in our parameter space of interest, $g_D \gg g_{Z^\prime}$, so we can neglect the SM contributions to the mediator width.

Furthermore, throughout our analysis, we will assume that $m_{Z^\prime} > m_\chi$ to ensure that all annihilation processes $\chi \bar\chi \to$ SM particles are $s$-channel and proceed through virtual $Z^\prime$ exchange. This requirement ensures that all annihilation and scattering rates depend on $g_{Z^\prime}$, either directly at tree level or indirectly through the kinetic mixing which also depends on this parameter.

\begin{figure}[t]
    \centering
    \includegraphics[width=0.7 \linewidth]{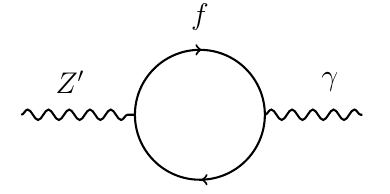}
    \caption{In each of the models we consider, there is an irreducible kinetic mixing that arises at loop level by integrating out all charged SM fermions $f$ that couple to both $\gamma$ and $Z^\prime$. In models where the $Z^\prime$ has tree-level couplings to electrons, this contribution is negligible. However, the $L_\mu - L_\tau$, $B-3L_\mu$, and  $B-3L_\tau$ mediators do not couple to electrons at tree level, so this induced kinetic mixing governs the $\chi$-$e$ direct-detection cross section.}
    \label{fig:kinmix}
\end{figure}

\subsection*{Mediator Particles}

Here we choose our mediator $Z^\prime$ from the list of all anomaly-free $U(1)$ gauge extensions to the SM, assuming there are no additional particles that participate in anomaly cancellation: 

 \medskip \noindent
 {\bfseries\itshape Dark Photon}:
    If SM particles are uncharged under the new $U(1)$ group, then the only allowed SM interaction is through the kinetic mixing interaction. In this scenario, $\varepsilon$ is taken to be a free parameter and $Z^\prime$ only couples to charged particles. Since this mediator is well-studied in the literature, we include it here only for completeness.

\medskip \noindent{\bfseries\itshape Gauged $\bm{ L_i - L_j}$}: For this class of models, two lepton families carry equal and opposite gauge charges to cancel triangle anomalies. The current from Eq.~\eqref{eq:Lint} can be written as
\begin{eqnarray}
    J_{Z^\prime}^\mu
     = \bar \ell_i\gamma^\mu \ell_i + \bar\nu_i\gamma^\mu P_L \nu_i - \bar \ell_j\gamma^\mu \ell_j - \bar\nu_j\gamma^\mu P_L\nu_j,
\end{eqnarray}
where $i\neq j = e,\mu,\tau$ are lepton family indices and $P_L = \frac{1}{2}(1-\gamma^5)$. The charged leptons $\ell_{i,j}$ couple to both the photon and the $Z^\prime$, so at loop level they induce a kinetic mixing between these vectors, as shown in Fig.~\ref{fig:kinmix} where the mixing parameter is 
\begin{eqnarray}
\label{eq:kin-mix-LiLj}
    \varepsilon = -\frac{e g_{Z'}}{12\pi^2}\log\frac{m^2_i}{m^2_j}\,.
\end{eqnarray}
For the $L_\mu-L_e$ and $L_e - L_\tau$ mediators, the kinetic mixing is a negligible correction to their tree-level electron coupling. However, for $L_\mu - L_\tau$, the electron coupling arises entirely from kinetic mixing, which governs the direct detection cross section (see Sec.~\ref{sec:constraints}).

 \medskip \noindent {\bfseries\itshape Gauged $\bm{B - L}$}: Here anomaly cancellation involves both the quarks and leptons. The current from Eq.~\eqref{eq:Lint} is 
\begin{eqnarray}
    J_{Z^\prime}^\mu = \frac{1}{3} \sum_q \bar q \gamma^\mu q - \sum_\ell (\bar \ell \gamma^\mu \ell  + \bar \nu_\ell \gamma^\mu \nu_\ell) ,
\end{eqnarray}
   and the induced kinetic mixing is 
    \begin{eqnarray}
    \label{eq:JBL}
         \varepsilon 
    = -\frac{{e g_{Z'}}}{18\pi^2}\log\left(\frac{m_d m_s m_b}{m_u^2 m_c^2 m_t^2}\frac{\Lambda^{12}}{m_e^3m_\mu^3 m_\tau^3}\right),
    \end{eqnarray}
    where $\Lambda$ is the unknown high-energy scale at which the loop is induced, typically coinciding with the scale of symmetry breaking for the new $U(1)$ group; unlike the $L_i -L_j$ scenario, the logarithmic divergence does not cancel in this case.
    However, this gauge boson couples to the first generation, so the kinetic mixing contribution to scattering signals is negligible. Note that although right-handed neutrinos are necessary for anomaly cancellation in the $B-L$ scenario, they can acquire large Majorana masses after spontaneous symmetry breaking and decouple; thus, we neglect them in Eq.~\eqref{eq:JBL}.

\medskip \noindent {\bfseries\itshape Gauged $\bm{B - 3L_i}$}: Here anomaly cancellation involves both quarks and leptons. The current from E 
q.~\eqref{eq:Lint} is 
\begin{eqnarray}
    J_{Z^\prime}^\mu
    = \frac{1}{3} \sum_q \bar q \gamma^\mu q-  3(\bar \ell_i \gamma^\mu \ell_i  + \bar \nu_{\ell_i} \gamma^\mu \nu_{\ell_i}) ,
\end{eqnarray}
and the loop-induced kinetic mixing  is
\begin{eqnarray}
    \varepsilon 
    = -\frac{{e g_{Z'}}}{18\pi^2}\log\left(\frac{m_d m_s m_b}{m_u^2 m_c^2 m_t^2}\frac{\Lambda^{12}}{m_{\ell_i}^9}\right).
\label{eq:epsB3L}
\end{eqnarray}
As in the $B-L$ case, the $B-3L_e$ mediator couples to electrons at tree-level, so the kinetic mixing is negligible. However, for $B-3L_{\mu,\tau}$ this mixing yields the only coupling to electrons and thereby governs direct detection signals. Since the scale $\Lambda$ depends on unspecified high-energy physics, we will consider a wide range between $100\,\text{GeV}$ and $M_{\rm Pl}$ as the theoretical uncertainty on the size of the kinetic mixing.  As in the $B-L$ scenario above, right-handed neutrinos are present at high energies for anomaly cancellation, but are decoupled in the infrared.

\begin{figure}[t!]
    \centering
    \includegraphics[width=0.48\textwidth]{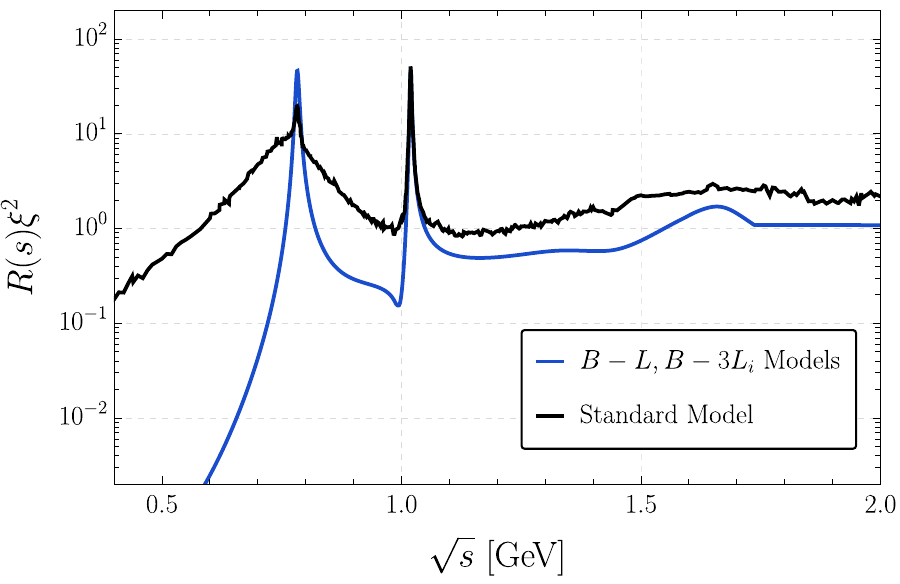}
    \caption{Normalized R-ratios for mediators with  tree-level quark couplings. The solid black curve shows SM R-ratio data~\cite{ParticleDataGroup:2020ssz} and corresponds to the choice $\xi = 1$; the dark photon mediator has the same R-ratio.
    The solid blue curve is the R-ratio for our other hadronically coupled mediators, with $\xi=Q^\prime_\mu$ for the $B-L$ and $B-3L_{\mu}$ extensions; for $B-3L_{e,\tau}$ there is no tree-level muon coupling, but there is an induced coupling through the kinetic mixing in Eq.~\eqref{eq:epsB3L}, so $\xi = \varepsilon Q_\mu$. }
    \label{fig:Rratio}
\end{figure}

\section{Cosmological Evolution}
\label{sec:relic}

In the early universe, $\chi$ is in equilibrium with the SM and the total DM number
density $n_\chi$ (including antiparticles) evolves as
\begin{equation}
\label{eq:boltz}
    \dot{n}_\chi + 3Hn_\chi = - \frac{1}{2}\langle \sigma v \rangle_{\rm ann} \left[n_\chi^2 - \left(n_\chi^{\text{eq}}\right)^2\right],
\end{equation}
where $H$ is the Hubble rate, $n_\chi^{\text{eq}}$ is the $\chi$ number density in chemical equilibrium with the SM, where we have assumed that the annihilating particles are not identical since only complex scalars and Dirac fermions are considered. 
To hit the DM relic density, starting from $n_\chi = n_\chi^{\text{eq}}$ at early times, the observed relic density is achieved for $\langle \sigma v \rangle_{\rm ann} \sim 10^{-26}\,\text{cm}^3\text{s}^{-1}$, with the precise value depending on mass and spin, which introduce variations of order unity~\cite{Steigman_2012}.

The thermally-averaged cross section in Eq.~\eqref{eq:boltz} can be written as~\cite{GONDOLO1991145}
\begin{eqnarray}
    \langle \sigma v \rangle_{\rm ann} = \frac{1}{N}\int_{4m_\chi^2}^\infty ds\,\sigma_{\rm ann}(s)\sqrt{s}(s-4m_\chi^2)K_1\left(\frac{\sqrt{s}}{T}\right),~~~
\label{eq:original_thermal_avg_cross_section}
\end{eqnarray}
where $\sigma_{\rm ann}$ is the total DM annihilation cross section, $s$ is the Mandelstam variable, $K_n$ is the n$^\text{th}$ order modified Bessel function,  and $N = 8m_\chi^4 T K_2^2(\frac{m_\chi}{T})$ is a normalization factor. 

In predictive models with $m_{Z^\prime} > m_\chi$, the interaction rate is mainly controlled by $s$-channel annihilation into SM fermions. For complex scalar DM, the cross section for DM annihilation into electromagnetically charged fermion pairs, $\chi \chi^* \to \bar f f$, is
\begin{equation}
    \sigma_{\chi\chi^*\to f\bar f} = \frac{4\pi\kappa_f\alpha_D(s+2m_f^2)\beta_\chi \beta_f}{3\left[(m_{Z'}^2-s)^2+m_{Z'}^2\Gamma_{Z'}^2\right]},
\end{equation}
where $\Gamma_{Z^\prime}$ is the $Z^\prime \to \chi \chi^*$ width from Eq.~\eqref{eq:WidthScalar},
$\beta_i=\sqrt{1-4m_i^2/s}$,
 and we have defined 
\begin{equation}
    \kappa_f =
    \begin{cases}
     Q_f^2 \, \varepsilon^2 \alpha, & \text{dark photon,} \\
     Q_f'^2 \alpha_{Z^\prime}, & \text{otherwise,}
    \end{cases}
\label{eq:kappaf}
\end{equation}
where $Q_f$ and $Q^\prime_{f}$ are respectively the electromagnetic and new  $U(1)$ charges of SM fermion $f$, and $\alpha_{Z^\prime} = g_{Z'}^2/4\pi$.

For Dirac fermion DM, the cross section per channel is
\begin{equation}
    \sigma_{\chi\bar\chi\to f\bar f} = \frac{4\pi\kappa_f\alpha_D(s+2m_\chi^2)(s+2m_f^2)}{3s\left[(m_{Z'}^2-s)^2+m_{Z'}^2
  \Gamma_{Z'}^2\right]}\frac{\beta_f}{\beta_\chi},
\end{equation}
where the decay width for $Z^\prime \to \chi \bar \chi$ is given in Eq.~\eqref{eq:WidthDirac}.  

For both complex scalar and Dirac fermion DM, the neutrino annihilation cross sections satisfies  
\begin{eqnarray}
\sigma_{\chi\bar\chi\to \nu \bar \nu} = \frac{1}{2}\sigma_{\chi\bar\chi\to f\bar f},\quad \kappa_\nu = Q_\nu'^2 \alpha_{Z^\prime}~, 
\end{eqnarray}
which is valid for all mediators considered here, except the dark photon, which does not couple to neutrinos.

\begin{figure}[t]
  \centering
  \includegraphics[width=0.47\textwidth]{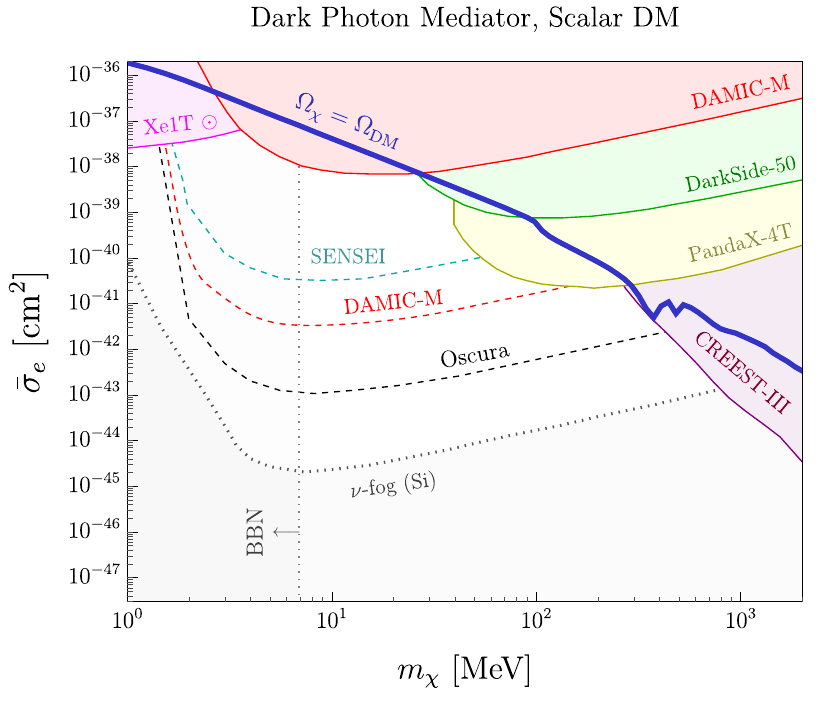}
  \caption{Limits on the effective DM-electron scattering cross section $\bar\sigma_e$ plotted against thermal relic targets for complex scalar DM in the dark photon model (adapted from Ref.~\cite{krnjaic2025testingthermalrelicdarkmatter}). Note that the Dirac fermion variant of this model is excluded by CMB data and is not shown here. Since the mediator is assumed to be heavier than the DM here, we use the $F_{\rm DM} = 1$ form factor~\cite{Lin:2019uvt}.}
  \label{fig:DarkPhotonScalar}
\end{figure}

 \medskip \noindent
{\bfseries\itshape Dark Photon}: For this model, the DM can annihilate into all kinematically allowed charged particles, including hadrons, so the total cross section is
\begin{eqnarray}
    \sigma_{\rm ann} = \sum_\ell
     \sigma_{\chi \bar\chi \to \ell \bar\ell}
     + \sigma_{\chi\bar \chi \to \text{had}}~,
\end{eqnarray}
and the cross section to hadrons is
\begin{equation}
\label{eq:R}
    \sigma_{\chi\bar \chi \to \text{had}} (s) = R
    (s)\,\sigma_{\chi \bar\chi \to \mu^+ \mu^-}(s),
\end{equation}
where $R = \sigma_{e^+e^- \to \rm had}/\sigma_{e^+e^- \to \mu^+\mu^-}$  is the  R-ratio function shown in Fig.~\ref{fig:Rratio} with data from Ref.~\cite{ParticleDataGroup:2024cfk}. 

 \medskip \noindent {\bfseries\itshape Gauged $\bm{L_i-L_j}$}: For the purely leptonic mediators 
\begin{equation}
    \sigma_\text{ann} = \!\!
   \sum_{\ell \,= \,\ell_i,\ell_j} 
   (\sigma_{\chi \bar\chi \to \ell \bar \ell} + 
   \sigma_{\chi \bar\chi \to \nu_\ell \bar\nu_\ell}),
\end{equation}
where hadronic annihilation channels are sharply suppressed as they only arise from the induced kinetic mixing in Eq.~\eqref{eq:kin-mix-LiLj}.

\medskip \noindent {\bfseries\itshape Gauged $\bm{B-L}$}: The annihilation cross section is
\begin{equation}
    ~~~~~\sigma_\text{ann} =
     \sum_{\ell} ( \sigma_{\chi \bar\chi \to \ell \bar \ell}  + \sigma_{\chi \bar\chi \to \nu_\ell \bar\nu_\ell}) + \sigma_{\chi\bar \chi \to \text{had}},
\end{equation}
where $\sigma_{\chi\bar \chi \to \text{had}}$ is defined in analogy with Eq.~\eqref{eq:R}, though the function $R(s)$ in this model differs from that of the dark photon case and we model this function following the prescription in Ref.~\cite{Foguel_2022}. For this mediator (and for all $B-3L_i$ variants), the normalized R-ratio is shown in Fig.~\ref{fig:Rratio}. 

\medskip \noindent {\bfseries\itshape Gauged $\bm{B-3L_i}$}:
Here the cross section is
\begin{align}
   ~~~~~\sigma_\text{ann} =   \sigma_{\chi \bar\chi \to \ell_i \bar \ell_i}  + \sigma_{\chi \bar\chi \to \nu_{\ell_i} \bar\nu_{\ell_i}} + \sigma_{\chi\bar \chi \to \text{had}},
\end{align}
where $\sigma_{\chi \bar \chi \to \rm had}$ is also defined in analogy with Eq.~\eqref{eq:R}, but with $R(s)$ for this model following Ref.~\cite{Foguel_2022}. 

\begin{figure*}[t]
  \centering
  \includegraphics[width=0.47\textwidth]{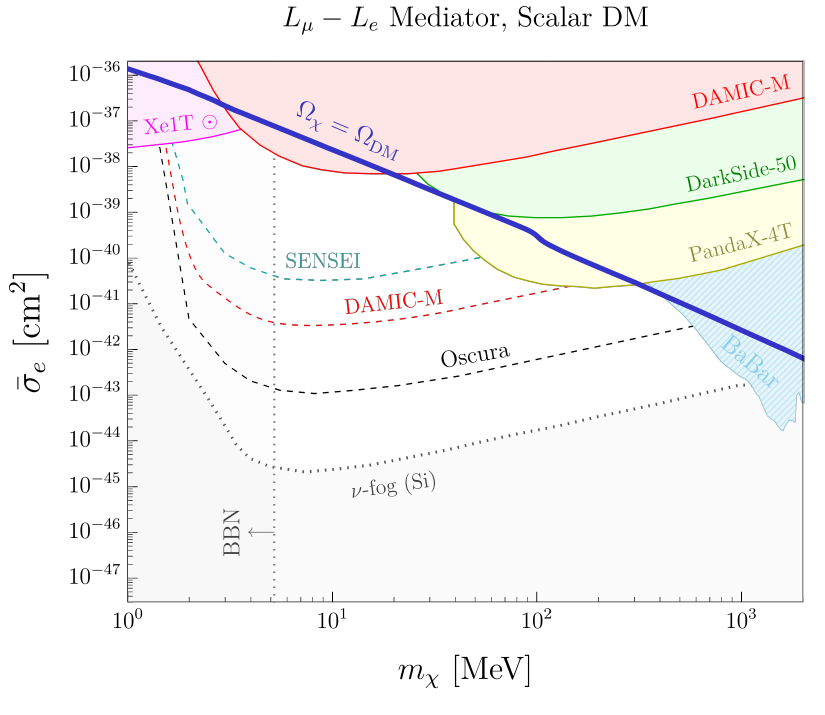}~~~
  \includegraphics[width=0.47\textwidth]{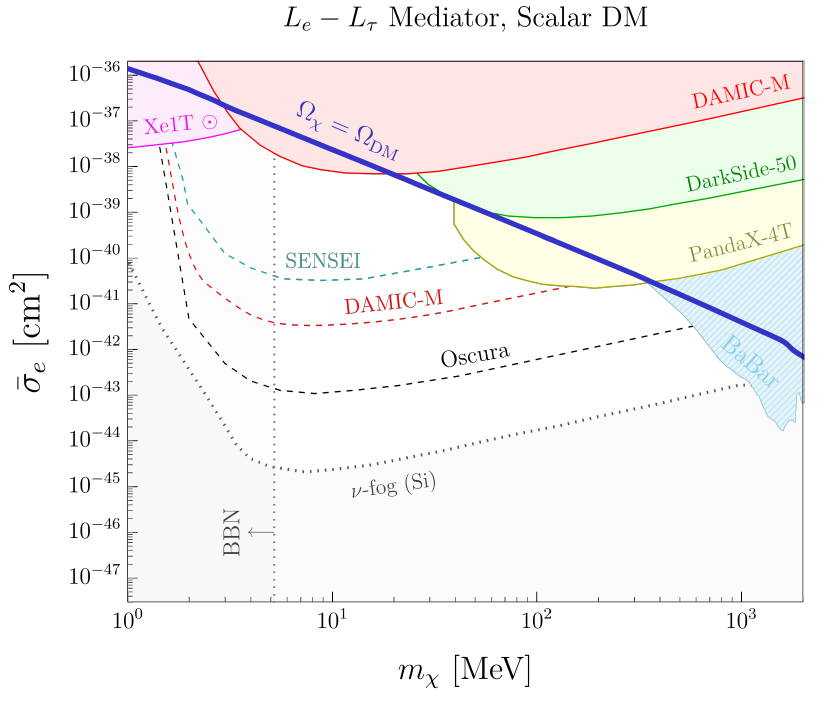}\\
    \includegraphics[width=0.47\textwidth]{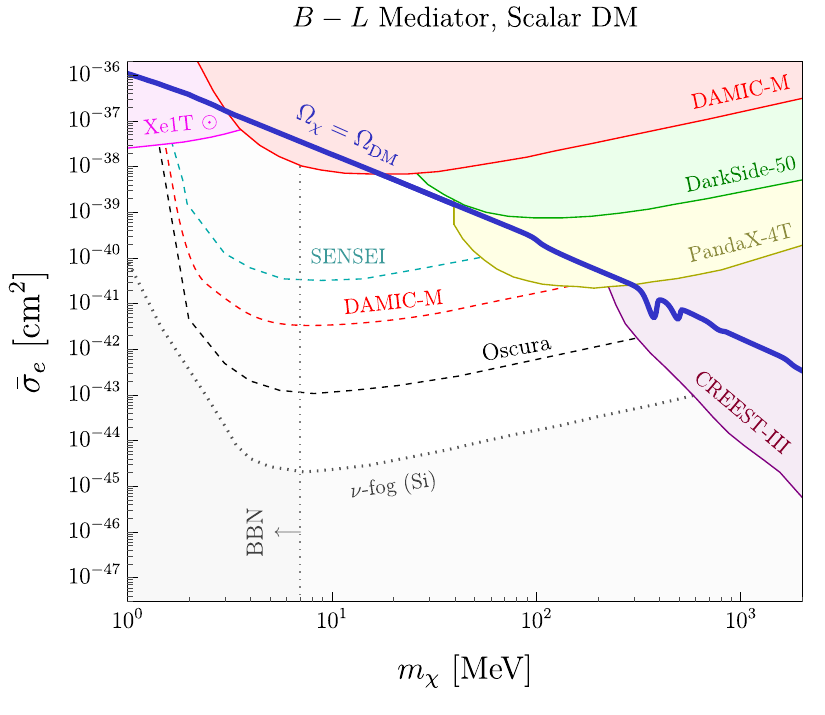}~~~
  \includegraphics[width=0.47\textwidth]{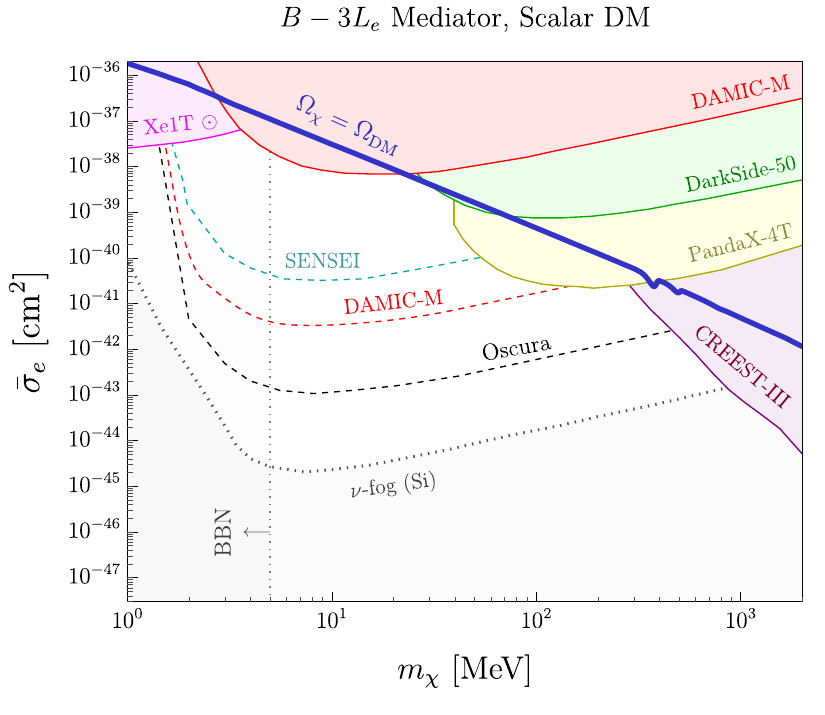}
  \caption{Thermal relic milestones for complex scalar DM coupled to the following mediators: $L_\mu-L_e$ (top left), $L_e-L_\tau$ (top right), $B-L$ (bottom left), and $B-3L_e$ (bottom right). Note that the Dirac fermion variants of these models are excluded by CMB energy injection limits (see Sec.~\ref{sec:constraints}). Since each of these mediators has a tree-level coupling to electrons, the parameter space is tightly constrained by existing limits, and resembles that of Fig.~\ref{fig:DarkPhotonScalar} and here we also work in the $F_{\rm DM} = 1$ regime~\cite{Lin:2019uvt}. Note that the cyan-shaded BABAR region (top row) is hatched to indicate that we chose $\alpha_D = 0.5$ and $m_{Z'} = 3 m_\chi$ to put this constraint on the $\bar \sigma_e$-$m_\chi$ plane; this choice is conservative in that smaller $\alpha_D$ or larger $m_{Z'}/m_\chi$ ratios would only make the constraint more severe~\cite{Izaguirre:2015yja}; no such assumption is required for the direct detection limits since these have the same parametric dependence as the annihilation cross section. The bottom region is the neutrino fog for electron scattering off a silicon target~\cite{Carew:2023qrj}.}
  \label{fig:electron-coupled}
\end{figure*}

\begin{figure*}[t]
  \centering
  \includegraphics[width=0.47\textwidth]{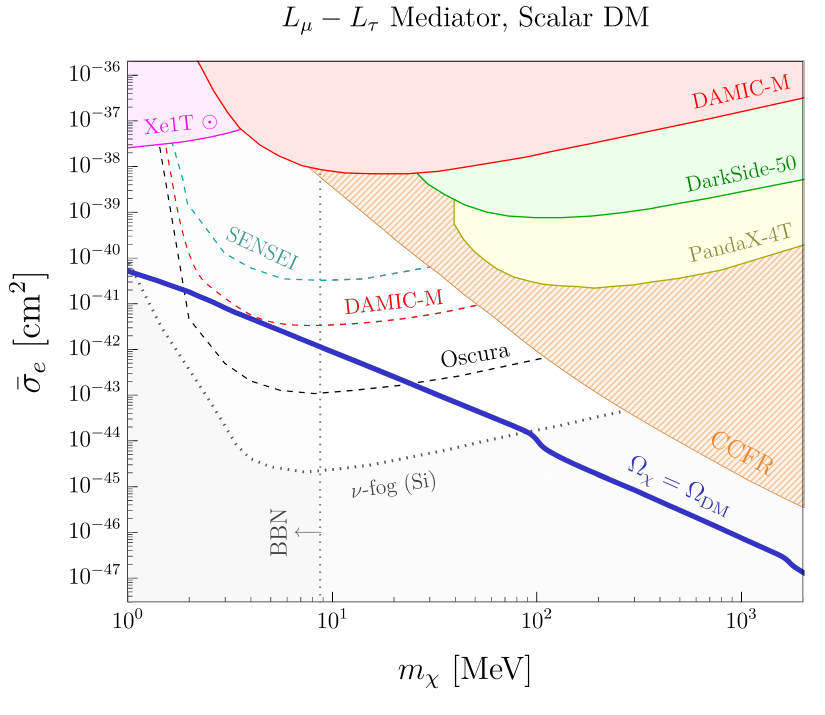}~~~
    \includegraphics[width=0.47\textwidth]{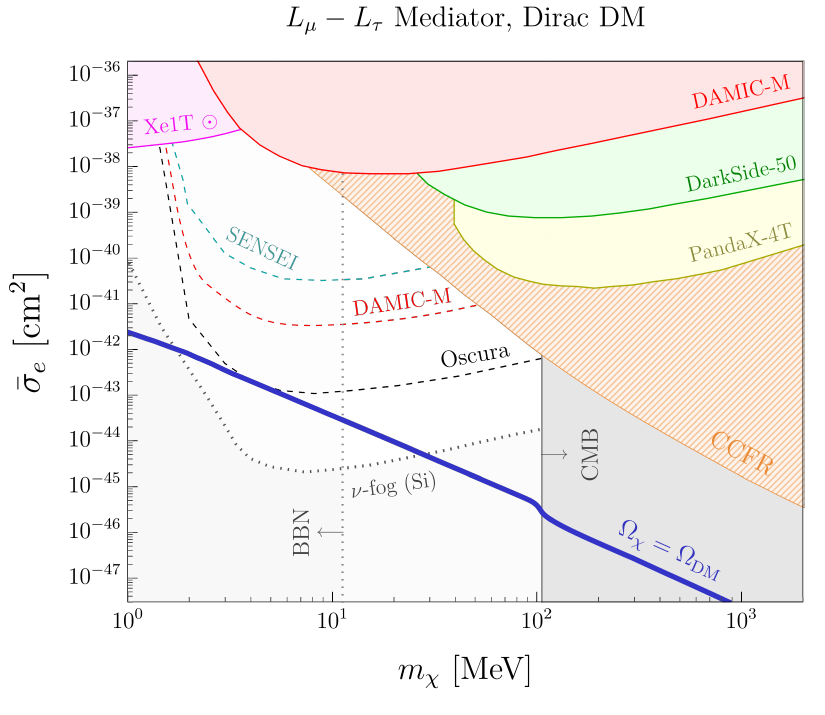}
  \caption{Limits on the effective DM-electron scattering cross section $\bar\sigma_e$ plotted against thermal relic targets for complex scalar (left) and Dirac fermion DM (right) in the $L_\mu-L_\tau$ model. Since the mediator here is heavier than the DM, we work in the $F_{\rm DM} = 1$ regime~\cite{Lin:2019uvt}.
  As in Fig.~\ref{fig:electron-coupled}, the orange CCFR region is hatched to indicate that we chose $\alpha_D = 0.5$ and $m_{Z'} = 3 m_\chi$ to put this constraint on the $\bar \sigma_e$-$m_\chi$ plane; this choice is conservative in that smaller $\alpha_D$ or larger $m_{Z'}/m_\chi$ ratios would only make the constraint more severe~\cite{Izaguirre:2015yja}. }
  \label{fig:LmuLtauScalar}
\end{figure*}

\begin{figure*}[t]
  \centering
  \includegraphics[width=0.47\textwidth]{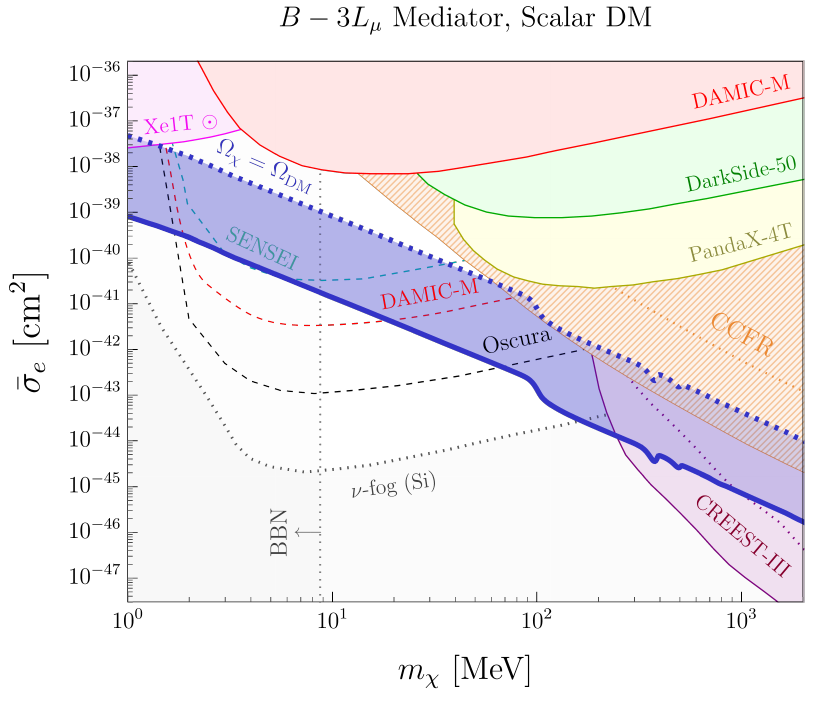}~~~
  \includegraphics[width=0.47\textwidth]{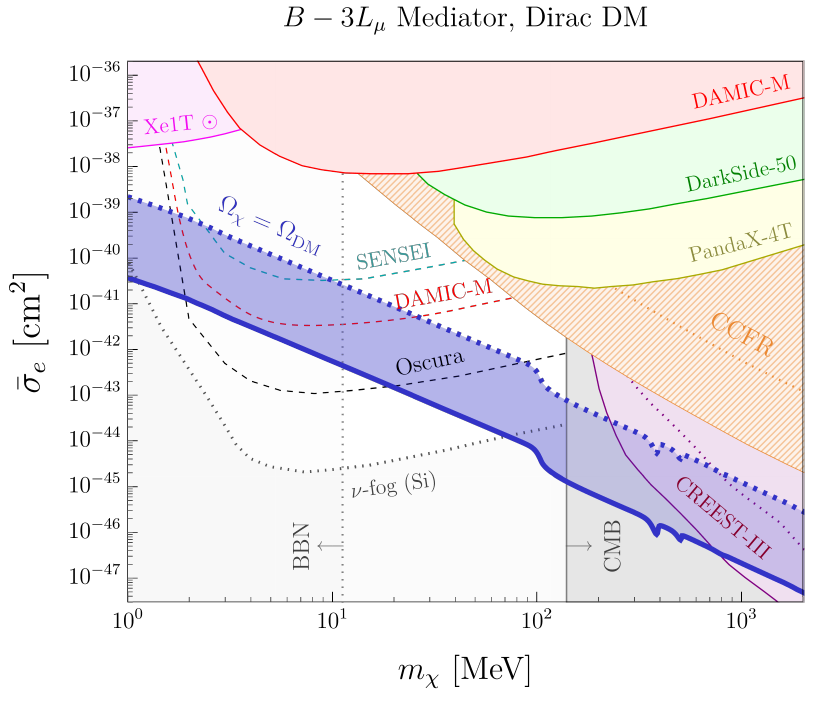} \\
    \includegraphics[width=0.47\textwidth]{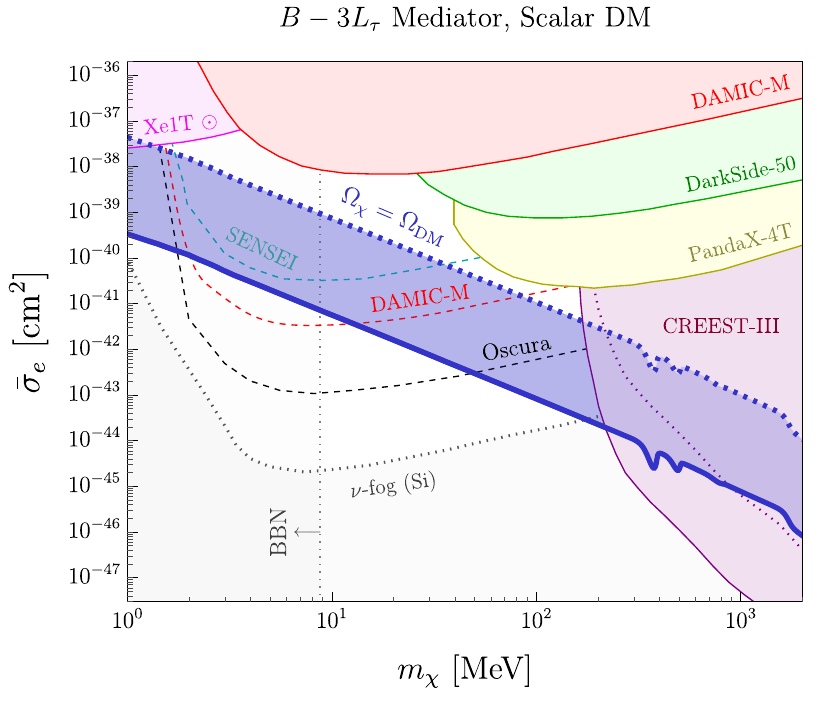}~~~
    \includegraphics[width=0.47\textwidth]{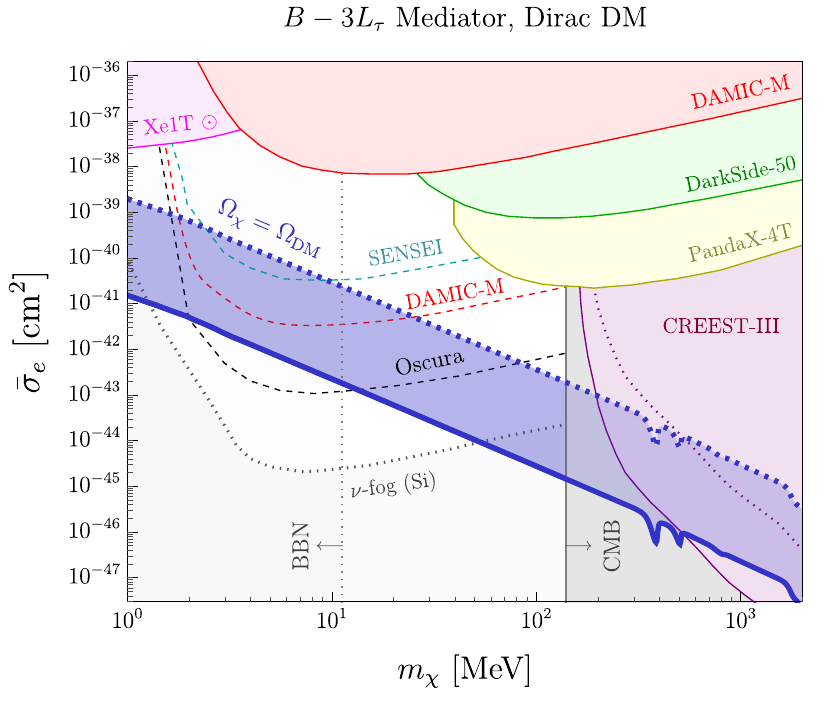}
  \caption{Top row: limits on the effective DM-electron scattering cross section $\bar\sigma_e$ plotted against thermal relic targets for complex scalar (left) and Dirac fermion DM (right) in the $B-3L_\mu$ mediator. Bottom row: same as the top, but for a $B-3L_\tau$ mediator. Here the annihilation cross section is independent of $\varepsilon$ at leading order, but translating the thermal-relic parameter space into an electron cross section requires introduces $\varepsilon$ dependence. Thus, unlike the other result plots in this paper, here the thermal relic target has a finite thickness due to the ultraviolet scale dependence in $\varepsilon$ from Eq.~\eqref{eq:epsB3L}. To remain agnostic about this scale, we vary $\Lambda$ between 100 GeV (bottom edge, solid curve) and $M_{\rm Pl}$  (upper edge, dotted curve). These choices also affect the CRESST-III~\cite{cresstcollaboration2024observationsinglephotonscresst} constraint   in this parameter space since the experiment constrains the {\it nucleon} cross section, which does not have this scale ambiguity, but translating these limits from $\sigma_N \to \bar \sigma_e$ requires an insertion of $\varepsilon^2$, which does; similarly the CCFR limits in the top row exhibit the same ambiguity since the muon coupling that the experiment constrains is not dependent on $\varepsilon$. For the CCFR and CRESST-III constraints in this plane, we use solid curves for $\Lambda = 100$ GeV  and dotted curves $\Lambda = M_{\rm Pl}$. }
  \label{fig:B3LtauScalar}
\end{figure*}

In our numerical results, we use $\langle\sigma v \rangle_{\rm ann}$ from Eq.~\eqref{eq:original_thermal_avg_cross_section} to identify the parameter space for the observed relic density following Ref.~\cite{Kolb:1990vq}. In  Figs.~\ref{fig:DarkPhotonScalar}-\ref{fig:B3LtauScalar}, we present the thermal-relic targets for these models in terms of the effective electron cross section (discussed in Sec.~\ref{sec:constraints}).

\section{Experimental Searches}
\label{sec:constraints}

\noindent{\bfseries\itshape Electron Recoils}:
Electron-recoil direct detection experiments offer a powerful probe of the models studied here because the scattering cross section depends on the same model parameters as the annihilation cross section. 
Conveniently, the scattering cross sections for scalar and Dirac fermions coupled to vector mediators have the same parametric form. We divide the mediators into two categories based on their couplings to electrons: 
\begin{itemize}
    \item  {\bfseries\itshape Electrophilic}: The dark photon, $L_\mu-L_e$, $L_e-L_\tau$, $B-L$,  and $B-3L_e$ mediators all couple to electrons at tree level, so the cross section for all of these can be written as
    \begin{equation}
        \bar\sigma_e = \frac{16\pi \kappa_e \alpha_D\mu^2_{\chi e}}{m_{Z'}^4}~,    \label{eq:ElectronRecoilCrossSectionElectrophilic}
    \end{equation}
    where $\kappa_e$ is given in Eq.~\eqref{eq:kappaf} and $\mu_{\chi e}$ is the $\chi$-$e$ reduced mass.
    \item  {\bfseries\itshape Electrophobic}: The $L_\mu - L_\tau$, $B-3L_\mu$, and $B-3L_\tau$ mediators have no tree-level couplings to electrons, so their leading-order cross section is
    \begin{equation}
        \bar\sigma_e = \frac{16\pi \varepsilon^2\alpha \alpha_D\mu^2_{\chi e}}{m_{Z'}^4}~,    \label{eq:ElectronRecoilCrossSectionElectrophobic}
    \end{equation}
    where $\varepsilon$ arises at one loop (as shown in Fig.~\ref{fig:kinmix}) and is given respectively in Eqs.~\eqref{eq:kin-mix-LiLj}, \eqref{eq:JBL}, and \eqref{eq:epsB3L}.
\end{itemize}
Since $m_{Z'} > m_\chi$ in the predictive parameter space, we trivially satisfy $m_{Z'} \gg \alpha m_e$, so the DM form factor is always $F_\text{DM}(q) = 1$ in our analysis~\cite{Lin:2019uvt}.

The DAMIC-M experiment has recently placed new bounds on $\bar \sigma_e$ with a $1.3 \,\text{kg·day}$ exposure on a Si target~\cite{DAMIC-M:2025luv}. The PandaX-4T has corresponding limits using an Xe target with a $0.55 \,\text{ton·year}$ exposure~\cite{Li_2023}. DarkSide-50  has reported $\bar\sigma_e$ bounds with
a $12,306 \,\text{kg·day}$ exposure on a liquid Ar target~\cite{Agnes_2023}; all of these limits are shown in Figs.~\ref{fig:DarkPhotonScalar}-\ref{fig:B3LtauScalar}. Related low-threshold direct-detection searches by SuperCDMS and EDELWEISS provide complementary constraints on light dark matter in nearby regions of parameter space, although we do not show these limits in our figures~\cite{SuperCDMS:2018mne,EDELWEISS:2019vjv}. These figures also show future sensitivity projections for the SENSEI~\cite{PhysRevLett.122.161801,SENSEI:2023gie}, DAMIC-M~\cite{Castell_Mor_2020}, and Oscura~\cite{aguilararevalo2022oscuraexperiment} experiments. The electron-DM cross section can also be probed using the sub-population of DM particles that are accelerated by the sun 
and can therefore deposit more energy into detectors~\cite{An_2021}; this constraint using XENON1T data \cite{XENON:2017vdw,XENON:2024xur} is shown in Figs.~\ref{fig:DarkPhotonScalar}-\ref{fig:B3LtauScalar}. 

\medskip

\noindent{\bfseries\itshape Nucleon Recoils}:
Our models of interest are also constrained by sub-GeV nuclear-recoil direct detection experiments. The cross sections for $\chi$-nucleon scattering are similar to those in Eqs.~\eqref{eq:ElectronRecoilCrossSectionElectrophilic} and \eqref{eq:ElectronRecoilCrossSectionElectrophobic} for electrophilic and electrophobic models, respectively. The only differences in the cross section formulas are that, for nucleon scattering,  we replace $\kappa_e \to \kappa_N$ where $N = p,n$ and $\mu_{\chi e} \to \mu_{\chi N}$, where the latter is the $\chi$-$N$ reduced mass.

In Figs.~\ref{fig:DarkPhotonScalar}-\ref{fig:B3LtauScalar}, for baryon-coupled mediators (dark photons, $B-L$, and $B-3L_i$) we show CRESST-III limits from Ref.~\cite{Abdelhameed_2019,cresstcollaboration2024observationsinglephotonscresst} which are based on a $3.64\, \text{kg·day}$ exposure with a $\text{CaWO}_4$ target. Note that limits in the CRESST-III experimental papers are plotted assuming DM coupled to all the nucleons inside the target. However, the dark photon mediator only couples to protons through kinetic mixing, so we account for the proton-nucleon ratio of $\text{CaWO}_4$ when translating limits on the nucleon cross section $\sigma_N$ into $\bar\sigma_e$.

The Migdal effect is also a powerful technique for extending DM-nucleon scattering sensitivity down to sub-GeV masses \cite{Ibe:2017yqa,Dolan:2017xbu,Baxter:2019pnz} and there are many reported experimental limits based on this signal contribution \cite{DarkSide:2022dhx,PandaX:2023xgl,XENON:2019zpr,CDEX:2019hzn,COSINE-100:2021poy,Armengaud_2022}. However, to-date the Migdal effect has not been robustly calibrated and there are conflicting reports about whether it has been observed in neutron scattering  \cite{Xu:2023wev,Yi:2026fmf}. Since detector-specific calibrations of the Migdal effect remain under development \cite{MIGDAL:2022yip,Xu:2026acq,Adams:2022zvg}, we do not include DM–nucleon scattering limits based on this effect. Should robust calibrations become available, these limits may provide an important test of the models considered here.

\medskip

\noindent{\bfseries\itshape B-Factories}:
The BABAR $e^+e^-$ experiment has been used to constrain $Z'$ production through the $e^+e^-\to Z' \gamma$ channel where the $Z'$ decays invisibly~\cite{Essig_2013,Izaguirre_2013}. The currently-running Belle-II experiment will eventually collect  $\sim 50\,\text{ab}^{-1}$ and is expected to greatly improve sensitivity to these particles.   Importantly, since the production cross section only depends on the SM coupling, placing $B$-factory constraints on our results plots requires a choice of $\alpha_D$ and $m_{Z^\prime}/m_\chi.$

\medskip
\noindent{\bfseries\itshape Neutrino Experiments}:
Neutrino fixed-target experiments can also probe new forces that interact with neutrinos and muons. The CCFR collaboration constrains such $Z'$ couplings via muon production through the trident process $\nu_\mu N \to \nu_\mu N \mu^+ \mu^-$~\cite{CCFR:1991lpl,Altmannshofer:2014pba}. In the future, the DUNE experiment is also poised to improve trident sensitivity to better probe new physics~\cite{Altmannshofer:2019zhy}.

\medskip
\noindent{\bfseries\itshape CMB}:
Light DM with an $s$-wave annihilation cross section faces stringent constraints from cosmic microwave background (CMB) energy injection near the epoch of recombination~\cite{Chen:2003gz,Padmanabhan:2005es,Galli:2009zc,Finkbeiner_2012}. To maintain consistency with Planck data, the cross section for annihilating to charged final states must satisfy~\cite{2020}
\begin{equation}
    \langle \sigma v \rangle_{\chi \bar\chi \to \bar ff}  \lesssim  2 \times 10^{-26} \, {\rm cm^3 s^{-1}}
      f^{-1}_{\rm eff}  \brac{m_\chi}{ 20 \rm \, GeV} \! ,
\label{eq:CMBConstriant}
\end{equation}
where $f_\text{\rm eff}$ parametrizes the ionization efficiency for a given final state and varies roughly between $0.1 - 1$, depending on the channel~\cite{Slatyer:2015jla}. Thus, for the Dirac models shown in Figs.~\ref{fig:LmuLtauScalar}-\ref{fig:B3LtauScalar} (right panels), the CMB rules out all sub-GeV thermal-relic parameter space unless the annihilation is predominantly to neutrinos. For the $L_\mu-L_\tau$, $B-3L_\mu$ and $B-3L_\tau$ mediators, Dirac fermion DM can still be viable for $\chi$ masses below the lightest charged particle to which the mediator couples. Since all scalar models feature $p$-wave annihilation, CMB energy injection does not meaningfully constrain their thermal-relic parameter space.

\begin{table*}[t!]
\centering
\small
\setlength{\tabcolsep}{5pt}
\begin{tabular}{lccccc}
\hline\hline
Mediator & Tree $e$ coupling & Dominant annihilation &   \hspace{-0.5cm} Dirac CMB status & $\varepsilon$ origin \\
\hline
Dark photon
    & Yes 
    & $\ell^+ \ell^-$, hadrons 
    &   Excluded 
    & Free parameter (dominant) \\

$L_\mu - L_e$ 
    & Yes 
    & $e^+e^-,\, \mu^+\mu^-,\bar \nu\nu$ 
    &    Excluded 
    & Loop-induced (negligible) \\

$L_e - L_\tau$ 
    & Yes 
    & $e^+e^-, \tau^+ \tau^-, \bar \nu\nu$ 
    &    Excluded 
    & Loop-induced (negligible) \\

$B - L$ 
    & Yes 
    & $\ell^+\ell^-, \bar \nu \nu,$ hadrons 
    &    Excluded 
    & Loop-induced (negligible) \\

$B - 3L_e$ 
    & Yes 
    & $e^+e^-, \bar \nu \nu$, hadrons 
    &    Excluded 
    & Loop-induced (negligible) \\

$L_\mu - L_\tau$ 
    & No 
    &  \hspace{-0.09cm}$\mu^+ \mu^-, \tau^+\tau^-,\nu\bar\nu$ ($\bar \nu\nu$ only for $m_\chi < m_\mu$) 
    &    Viable 
    & Loop-induced (dominant) \\

$B - 3L_\mu$ 
    & No 
    &  \hspace{-0.5cm} $\mu^+\mu^-,\nu\bar\nu$, hadrons ($\bar \nu \nu$ only for $m_\chi < m_\mu$) 
    &   Viable 
    & Loop-induced (dominant) \\

$B - 3L_\tau$ 
    & No 
    &  \hspace{-0.5cm} $\tau^+\tau^-,\nu\bar\nu$, hadrons ($\bar \nu \nu$ only for $m_\chi < m_\tau$) 
    &   Viable 
    & Loop-induced (dominant) \\
\hline\hline
\end{tabular}
\caption{Summary of anomaly-free $U(1)$ mediators considered in this work, assuming $m_{Z^\prime} > m_{\chi}$ so that there is a predictive thermal target. The fourth column summarizes whether Dirac fermion DM is excluded by the CMB for a given mediator; all complex scalar models here feature $p$-wave annihilation and are not constrained by CMB data. The fifth column lists whether the kinetic mixing parameter, $\varepsilon$, arises at loop level and also whether it provides the dominant electron-scattering process (relative to the tree-level coupling) for a given model, or just adds a negligible correction. }
\label{tab:mediator_summary}
\end{table*}


\medskip
\noindent{\bfseries\itshape BBN}:
For MeV scale with a thermal-relic abundance, freeze-out occurs around the epoch of big bang nucleosynthesis (BBN) and modifies the effective number of relativistic species at this time. Preserving the successful SM prediction of light element yields constrains on the mass scale of thermal DM candidates by requiring freeze-out to occur earlier before neutrinos decouple to set the initial conditions for BBN. In Figs.~\ref{fig:DarkPhotonScalar}-\ref{fig:B3LtauScalar} result plots, we show BBN bounds on $m_\chi$ from Refs.~\cite{Nollett_2014,B_hm_2013,Krnjaic_2020,Sabti_2020,Chu:2022qnb,Chu:2023fmc}.

\section{Results}
\label{sec:results}

In Figs.~\ref{fig:DarkPhotonScalar}-\ref{fig:B3LtauScalar} we present thermal relic parameter space for complex scalar and Dirac fermion DM candidates in terms of the effective electron cross section $\bar \sigma_e$ (blue curves). Each plot also shows a variety of constraints from direct detection experiments and accelerator searches, where applicable. We divide our results into two categories depending on whether a given model has tree-level couplings to electrons.

\medskip

\noindent {\bfseries\itshape Electrophilic Models}: In this category we include all models with the dark photon, $L_\mu-L_e$, $L_e -L_\tau$, $B-L$, and $B-3L_e$ mediators which couple appreciably to electrons. 

In Fig.~\ref{fig:DarkPhotonScalar} we take results from Ref.~\cite{krnjaic2025testingthermalrelicdarkmatter} which found that scalar DM coupled to a dark photon mediator is excluded across the entire sub-GeV parameter space in the predictive regime where $m_{Z^\prime} > m_\chi$ and the annihilation and direct detection cross sections depend on the same parameters. We do not show the Dirac fermion version of this model because the annihilation cross section is $s$-wave, so all sub-GeV parameter space is excluded by CMB limits (see Sec.~\ref{sec:constraints}).

In Fig.~\ref{fig:electron-coupled}, we show four panels with the parameter space for the $L_\mu-L_e$, $L_e -L_\tau$, $B-L$, and $B-3L_e$ mediators coupled to complex scalar DM. As in the dark photon case, the parameter space is almost fully excluded by direct detection searches, except for small regions within the reach of existing experiments, which will fully test these milestones soon. As in the dark photon case, the Dirac fermion variants of these models are all excluded by CMB limit from Eq.~\eqref{eq:CMBConstriant}, so we do not show these plots. 

\medskip

\noindent {\bfseries\itshape Electrophobic Models}:
In this category we include all models coupled to the $L_\mu - L_\tau$, $B -3L_\mu$, and $B -3L_\tau$ mediators whose electron coupling only arises from the one-loop contribution shown in Fig.~\ref{fig:kinmix}. Importantly, for these models we include results for both complex scalar and Dirac fermion DM. Unlike electrophilic models, here in each case the dominant annihilation process is  $\chi \bar\chi \to \bar \nu \nu$ for $m_\chi$ below the lightest charged particle to which the mediator couples (e.g., $m_\chi < m_\mu$ for $L_\mu - L_\tau$). Since annihilation to neutrinos is not constrained by CMB energy injection limits, the Dirac models can be cosmologically viable below a certain mass threshold for these mediators.

In Fig.~\ref{fig:LmuLtauScalar} we show results for the $L_\mu -L_\tau$ model. In these plots the thermal-relic targets are significantly lower due to the kinetic mixing suppression from Eq.~\eqref{eq:kin-mix-LiLj}. Furthermore, this model is subject to limits on neutrino trident production  from CCFR~\cite{CCFR:1991lpl} (see Sec.~\ref{sec:constraints}).

Finally, in Fig.~\ref{fig:B3LtauScalar} we show parameter space for the $B-3L_{\mu,\tau}$ mediators. For these figures, the thermal relic region has a finite thickness due to the UV sensitivity of the kinetic mixing. Although the annihilation cross section is not UV sensitive, the direct detection cross section depends on the unknown parameter $\Lambda$ in Eq.~\eqref{eq:epsB3L}, which represents the energy scale at which the kinetic mixing from Fig.~\ref{fig:kinmix} is induced. Thus, to parametrize our ignorance, we allow this scale to vary between $100\,\text{GeV}$ (bottom of the relic band) and the Planck scale, $M_{\rm Pl} = 1.2 \times 10^{19}$ GeV (top of the relic band).

\section{Discussion}
\label{sec:discussion}

In this paper we have identified thermal-relic milestones for sub-GeV direct detection searches. Our focus has been on models of scalar and fermion DM coupled to mediators from anomaly-free $U(1)$ extensions to the SM. Furthermore, we have restricted ourselves to the $m_{Z'} > m_\chi$ parameter space for which the direct-detection cross section is in one-to-one correspondence with the thermal-relic cross section, which offers an experimental target for discovery or falsification. 

We have found that predictive models with tree-level electron couplings are either ruled out by existing constraints, or will soon be fully tested in the few remaining viable parameter regions. However, models with mediators do not couple to electrons at tree-level are broadly viable in the MeV-GeV mass range that is compatible with cosmological bounds and testable with low-mass direct detection experiments. Our core findings are summarized in Table \ref{tab:mediator_summary}. 
 
For each mediator, we have considered complex scalar and Dirac fermion DM for simplicity. While variations of these models with Majorana or pseudo-Dirac fermion DM are also viable, their direct-detection cross sections are either velocity- or further loop- suppressed  relative to the scenarios studied here. Thus, these variations are not very promising for low-mass direct detection at this time, though some promising strategies with crystal detectors may eventually succeed at reaching these milestones~\cite{Kahn:2020fef}. 

Throughout this paper we have emphasized direct detection signals and constraints because these depend on the same combination of model parameters as the annihilation cross section that sets the relic density. However, these models can also be probed via accelerator production at NA64~\cite{NA64:2025ddk}, LDMX~\cite{LDMX:2018cma,Izaguirre:2014bca}, BDX~\cite{Izaguirre_2013,BDX:2016akw}, DUNE~\cite{DeRomeri:2019kic}, MiniBooNE~\cite{MiniBooNEDM:2018cxm}, MUonE~\cite{krnjaic2025testingthermalrelicdarkmatter}, JSNS2~\cite{Jordan:2018gcd}, ATLAS~\cite{Galon:2019owl}, CMS~\cite{Izaguirre_2013,CMS:2023bay}, FASER~\cite{Dienes:2023uve,Berlin:2018jbm}, and Belle-II~\cite{Duerr:2019dmv}, among others (see Ref.~\cite{Krnjaic:2022ozp} for a review). For simplicity, we have not included projections for these experiments as accelerator signals depend only on the SM-mediator coupling $\alpha_{Z^\prime}$ and mass $m_{Z'}$, but  not on $\alpha_D$ or $m_\chi$. Thus, it is necessary to choose values for these quantities in order to place accelerator projections onto the $\bar \sigma_e$-$m_\chi$ parameter space.

Finally, we note that our list of minimal anomaly-free $U(1)$ gauge extensions is complete up to additional linear combinations of the possibilities considered here \cite{Foguel_2022}, assuming only SM field content (and right handed neutrinos in certain cases). However, other options might be possible if additional SM-charged states are added to these theories as ``anomalons." In these scenarios, there are additional constraints on the gauge couplings and mediator masses from LHC searches for the anomalons themselves \cite{Kahn:2016vjr,Dobrescu:2014fca,Dror:2017ehi,Dror:2017nsg}. Furthermore, it is also possible to gauge the anomaly-free combination  $B_i - B_j$ where $i$ indexes baryonic family number, however these models are generically face stringent constraints from flavor changing neutral currents \cite{Altmannshofer:2019xda}, which makes these mediators more challenging for our purposes here; we leave these scenarios for future work.

\medskip
{\bf Acknowledgments.}
We thank Dan Baxter and Paolo Privitera for helpful conversations. This manuscript has been authored in part by  Fermi Forward Discovery Group, LLC under Contract No. 89243024CSC000002 with the U.S. Department of Energy, Office of Science, Office of High Energy Physics. and by the Kavli Institute for Cosmological Physics at the University of Chicago through an
endowment from the Kavli Foundation and its founder
Fred Kavli. 

\bibliography{biblio}

@article{DAMIC-M:2025luv,
    author = {Aggarwal, K. and others},
    collaboration = {DAMIC-M},
    title = {{Probing Benchmark Models of Hidden-Sector Dark Matter with DAMIC-M}},
    eprint = {2503.14617},
    archivePrefix = {arXiv},
    primaryClass = {hep-ex},
    doi = {10.1103/2tcc-bqck},
    journal = {Phys. Rev. Lett.},
    volume = {135},
    number = {7},
    pages = {071002},
    year = {2025}
}

@article{ParticleDataGroup:2024cfk,
    author = {Navas, S. and others},
    collaboration = {Particle Data Group},
    title = {{Review of particle physics}},
    doi = {10.1103/PhysRevD.110.030001},
    journal = {Phys. Rev. D},
    volume = {110},
    number = {3},
    pages = {030001},
    year = {2024}
}

@misc{Cirelli:2024ssz,
    author = {Cirelli, Marco and Strumia, Alessandro and Zupan, Jure},
    title = {{Dark Matter}},
    eprint = {2406.01705},
    archivePrefix = {arXiv},
    primaryClass = {hep-ph},
    month = {6},
    year = {2024}
}

@article{Kahn:2020fef,
    author = {Kahn, Yonatan and Krnjaic, Gordan and Mandava, Bashi},
    title = {{Dark Matter Detection with Bound Nuclear Targets: The Poisson Phonon Tail}},
    eprint = {2011.09477},
    archivePrefix = {arXiv},
    primaryClass = {hep-ph},
    reportNumber = {FERMILAB-PUB-20-588-T},
    doi = {10.1103/PhysRevLett.127.081804},
    journal = {Phys. Rev. Lett.},
    volume = {127},
    number = {8},
    pages = {081804},
    year = {2021}
}

@misc{Krnjaic:2022ozp,
    author = {Krnjaic, G. and others},
    title = {{A Snowmass Whitepaper: Dark Matter Production at Intensity-Frontier Experiments}},
    eprint = {2207.00597},
    archivePrefix = {arXiv},
    primaryClass = {hep-ph},
    reportNumber = {FERMILAB-PUB-22-497-T},
    month = {7},
    year = {2022}
}

@article{Galon:2019owl,
    author = {Galon, Iftah and Kajamovitz, Enrique and Shih, David and Soreq, Yotam and Tarem, Shlomit},
    title = {{Searching for muonic forces with the ATLAS detector}},
    eprint = {1906.09272},
    archivePrefix = {arXiv},
    primaryClass = {hep-ph},
    reportNumber = {CERN-TH-2019-097},
    doi = {10.1103/PhysRevD.101.011701},
    journal = {Phys. Rev. D},
    volume = {101},
    number = {1},
    pages = {011701},
    year = {2020}
}

@article{Duerr:2019dmv,
    author = {Duerr, Michael and Ferber, Torben and Hearty, Christopher and Kahlhoefer, Felix and Schmidt-Hoberg, Kai and Tunney, Patrick},
    title = {{Invisible and displaced dark matter signatures at Belle II}},
    eprint = {1911.03176},
    archivePrefix = {arXiv},
    primaryClass = {hep-ph},
    reportNumber = {DESY-19-141, OUTP-19-10P, TTK-19-46},
    doi = {10.1007/JHEP02(2020)039},
    journal = {JHEP},
    volume = {02},
    pages = {039},
    year = {2020}
}

@article{Berlin:2018jbm,
    author = {Berlin, Asher and Kling, Felix},
    title = {{Inelastic Dark Matter at the LHC Lifetime Frontier: ATLAS, CMS, LHCb, CODEX-b, FASER, and MATHUSLA}},
    eprint = {1810.01879},
    archivePrefix = {arXiv},
    primaryClass = {hep-ph},
    doi = {10.1103/PhysRevD.99.015021},
    journal = {Phys. Rev. D},
    volume = {99},
    number = {1},
    pages = {015021},
    year = {2019}
}

@article{Dienes:2023uve,
    author = {Dienes, Keith R. and Feng, Jonathan L. and Fieg, Max and Huang, Fei and Lee, Seung J. and Thomas, Brooks},
    title = {{Extending the discovery potential for inelastic-dipole dark matter with FASER}},
    eprint = {2301.05252},
    archivePrefix = {arXiv},
    primaryClass = {hep-ph},
    reportNumber = {UCI-TR-2022-16},
    doi = {10.1103/PhysRevD.107.115006},
    journal = {Phys. Rev. D},
    volume = {107},
    number = {11},
    pages = {115006},
    year = {2023}
}

@article{CMS:2023bay,
    author = {Hayrapetyan, Aram and others},
    collaboration = {CMS},
    title = {{Search for Inelastic Dark Matter in Events with Two Displaced Muons and Missing Transverse Momentum in Proton-Proton Collisions at s=13{\,}{\,}TeV}},
    eprint = {2305.11649},
    archivePrefix = {arXiv},
    primaryClass = {hep-ex},
    reportNumber = {CMS-EXO-20-010, CERN-EP-2023-083},
    doi = {10.1103/PhysRevLett.132.041802},
    journal = {Phys. Rev. Lett.},
    volume = {132},
    number = {4},
    pages = {041802},
    year = {2024}
}

@article{Jordan:2018gcd,
    author = {Jordan, Johnathon R. and Kahn, Yonatan and Krnjaic, Gordan and Moschella, Matthew and Spitz, Joshua},
    title = {{Signatures of Pseudo-Dirac Dark Matter at High-Intensity Neutrino Experiments}},
    eprint = {1806.05185},
    archivePrefix = {arXiv},
    primaryClass = {hep-ph},
    reportNumber = {FERMILAB-PUB-18-148-A, PUPT 2563, PUPT-2563},
    doi = {10.1103/PhysRevD.98.075020},
    journal = {Phys. Rev. D},
    volume = {98},
    number = {7},
    pages = {075020},
    year = {2018}
}

@article{MiniBooNEDM:2018cxm,
    author = {Aguilar-Arevalo, A. A. and others},
    collaboration = {MiniBooNE DM},
    title = {{Dark Matter Search in Nucleon, Pion, and Electron Channels from a Proton Beam Dump with MiniBooNE}},
    eprint = {1807.06137},
    archivePrefix = {arXiv},
    primaryClass = {hep-ex},
    reportNumber = {LA-UR-18-26421, FERMILAB-PUB-18-334-ND},
    doi = {10.1103/PhysRevD.98.112004},
    journal = {Phys. Rev. D},
    volume = {98},
    number = {11},
    pages = {112004},
    year = {2018}
}

@article{DeRomeri:2019kic,
    author = {De Romeri, Valentina and Kelly, Kevin J. and Machado, Pedro A. N.},
    title = {{DUNE-PRISM Sensitivity to Light Dark Matter}},
    eprint = {1903.10505},
    archivePrefix = {arXiv},
    primaryClass = {hep-ph},
    reportNumber = {FERMILAB-PUB-19-116-T},
    doi = {10.1103/PhysRevD.100.095010},
    journal = {Phys. Rev. D},
    volume = {100},
    number = {9},
    pages = {095010},
    year = {2019}
}

@misc{BDX:2016akw,
    author = {Battaglieri, M. and others},
    collaboration = {BDX},
    title = {{Dark Matter Search in a Beam-Dump eXperiment (BDX) at Jefferson Lab}},
    eprint = {1607.01390},
    archivePrefix = {arXiv},
    primaryClass = {hep-ex},
    reportNumber = {FERMILAB-TM-2630-PPD},
    month = {7},
    year = {2016}
}

@article{Izaguirre:2014bca,
    author = {Izaguirre, Eder and Krnjaic, Gordan and Schuster, Philip and Toro, Natalia},
    title = {{Testing GeV-Scale Dark Matter with Fixed-Target Missing Momentum Experiments}},
    eprint = {1411.1404},
    archivePrefix = {arXiv},
    primaryClass = {hep-ph},
    doi = {10.1103/PhysRevD.91.094026},
    journal = {Phys. Rev. D},
    volume = {91},
    number = {9},
    pages = {094026},
    year = {2015}
}

@misc{LDMX:2018cma,
    author = {{\r{A}}kesson, Torsten and others},
    collaboration = {LDMX},
    title = {{Light Dark Matter eXperiment (LDMX)}},
    eprint = {1808.05219},
    archivePrefix = {arXiv},
    primaryClass = {hep-ex},
    reportNumber = {FERMILAB-PUB-18-324-A, SLAC-PUB-17303},
    month = {8},
    year = {2018}
}

@misc{NA64:2025ddk,
    author = {Andreev, Yu. M. and others},
    collaboration = {NA64},
    title = {{Searching for Light Dark Matter and Dark Sectors with the NA64 experiment at the CERN SPS}},
    eprint = {2505.14291},
    archivePrefix = {arXiv},
    primaryClass = {hep-ex},
    month = {5},
    year = {2025}
}

@article{Altmannshofer:2019zhy,
    author = {Altmannshofer, Wolfgang and Gori, Stefania and Mart{\'\i}n-Albo, Justo and Sousa, Alexandre and Wallbank, Michael},
    title = {{Neutrino Tridents at DUNE}},
    eprint = {1902.06765},
    archivePrefix = {arXiv},
    primaryClass = {hep-ph},
    reportNumber = {FERMILAB-PUB-19-062-LBNF-ND},
    doi = {10.1103/PhysRevD.100.115029},
    journal = {Phys. Rev. D},
    volume = {100},
    number = {11},
    pages = {115029},
    year = {2019}
}

@book{Kolb:1990vq,
    author = {Kolb, Edward W. and Turner, Michael S.},
    title = {{The Early Universe}},
    reportNumber = {FERMILAB-BOOK-1990-01},
    doi = {10.1201/9780429492860},
    isbn = {978-0-429-49286-0, 978-0-201-62674-2},
    publisher = {Taylor and Francis},
    volume = {69},
    month = {5},
    year = {2019}
}

@article{Feng:2017drg,
    author = {Feng, Jonathan L. and Smolinsky, Jordan},
    title = {{Impact of a resonance on thermal targets for invisible dark photon searches}},
    eprint = {1707.03835},
    archivePrefix = {arXiv},
    primaryClass = {hep-ph},
    reportNumber = {UCI-TR-2017-06},
    doi = {10.1103/PhysRevD.96.095022},
    journal = {Phys. Rev. D},
    volume = {96},
    number = {9},
    pages = {095022},
    year = {2017}
}

@article{Izaguirre:2015yja,
    author = {Izaguirre, Eder and Krnjaic, Gordan and Schuster, Philip and Toro, Natalia},
    title = {{Analyzing the Discovery Potential for Light Dark Matter}},
    eprint = {1505.00011},
    archivePrefix = {arXiv},
    primaryClass = {hep-ph},
    doi = {10.1103/PhysRevLett.115.251301},
    journal = {Phys. Rev. Lett.},
    volume = {115},
    number = {25},
    pages = {251301},
    year = {2015}
}

@article{Berlin:2018bsc,
    author = {Berlin, Asher and Blinov, Nikita and Krnjaic, Gordan and Schuster, Philip and Toro, Natalia},
    title = {{Dark Matter, Millicharges, Axion and Scalar Particles, Gauge Bosons, and Other New Physics with LDMX}},
    eprint = {1807.01730},
    archivePrefix = {arXiv},
    primaryClass = {hep-ph},
    reportNumber = {FERMILAB-PUB-18-310-A, SLAC-PUB-17297},
    doi = {10.1103/PhysRevD.99.075001},
    journal = {Phys. Rev. D},
    volume = {99},
    number = {7},
    pages = {075001},
    year = {2019}
}

@article{Bauer:2018onh,
    author = {Bauer, Martin and Foldenauer, Patrick and Jaeckel, Joerg},
    title = {{Hunting All the Hidden Photons}},
    eprint = {1803.05466},
    archivePrefix = {arXiv},
    primaryClass = {hep-ph},
    doi = {10.1007/JHEP07(2018)094},
    journal = {JHEP},
    volume = {07},
    pages = {094},
    year = {2018}
}

@article{Griest:1989wd,
    author = {Griest, Kim and Kamionkowski, Marc},
    title = {{Unitarity Limits on the Mass and Radius of Dark Matter Particles}},
    reportNumber = {CFPA-TH-89-013, FERMILAB-PUB-89-205-A},
    doi = {10.1103/PhysRevLett.64.615},
    journal = {Phys. Rev. Lett.},
    volume = {64},
    pages = {615},
    year = {1990}
}

@article{Lee:1977ua,
    author = {Lee, Benjamin W. and Weinberg, Steven},
    editor = {Srednicki, M. A.},
    title = {{Cosmological Lower Bound on Heavy Neutrino Masses}},
    reportNumber = {FERMILAB-PUB-77-041-T},
    doi = {10.1103/PhysRevLett.39.165},
    journal = {Phys. Rev. Lett.},
    volume = {39},
    pages = {165--168},
    year = {1977}
}

@article{Mitridate:2022tnv,
    author = {Mitridate, Andrea and Trickle, Tanner and Zhang, Zhengkang and Zurek, Kathryn M.},
    title = {{Snowmass white paper: Light dark matter direct detection at the interface with condensed matter physics}},
    eprint = {2203.07492},
    archivePrefix = {arXiv},
    primaryClass = {hep-ph},
    reportNumber = {FERMILAB-PUB-23-551-T},
    doi = {10.1016/j.dark.2023.101221},
    journal = {Phys. Dark Univ.},
    volume = {40},
    pages = {101221},
    year = {2023}
}

@article{Lin:2019uvt,
    author = {Lin, Tongyan},
    title = {{Dark matter models and direct detection}},
    eprint = {1904.07915},
    archivePrefix = {arXiv},
    primaryClass = {hep-ph},
    doi = {10.22323/1.333.0009},
    journal = {PoS},
    volume = {333},
    pages = {009},
    year = {2019}
}

@article{Foguel_2022,
    author = {Foguel, Ana Luisa and Reimitz, Peter and Funchal, Renata Zukanovich},
    title = {{A robust description of hadronic decays in light vector mediator models}},
    journal = {JHEP},
    volume = {04},
    pages = {119},
    year = {2022},
    doi = {10.1007/JHEP04(2022)119},
    eprint = {2201.01788},
    archivePrefix = {arXiv},
    primaryClass = {hep-ph}
}

@article{ParticleDataGroup:2020ssz,
    author = {Zyla, P. A. and others},
    collaboration = {Particle Data Group},
    title = {{Review of Particle Physics}},
    journal = {PTEP},
    volume = {2020},
    number = {8},
    pages = {083C01},
    year = {2020},
    doi = {10.1093/ptep/ptaa104}
}

@article{Slatyer:2015jla,
    author = {Slatyer, Tracy R.},
    title = {{Indirect dark matter signatures in the cosmic dark ages. I. Generalizing the bound on s-wave dark matter annihilation from Planck results}},
    journal = {Phys. Rev. D},
    volume = {93},
    number = {2},
    pages = {023527},
    year = {2016},
    doi = {10.1103/PhysRevD.93.023527},
    eprint = {1506.03811},
    archivePrefix = {arXiv},
    primaryClass = {hep-ph}
}

@article{PhysRevLett.122.161801,
    author = {Abramoff, Orr and others},
    collaboration = {SENSEI},
    title = {{SENSEI: Direct-Detection Constraints on Sub-GeV Dark Matter from a Shallow Underground Run Using a Prototype Skipper CCD}},
    journal = {Phys. Rev. Lett.},
    volume = {122},
    number = {16},
    pages = {161801},
    year = {2019},
    doi = {10.1103/PhysRevLett.122.161801},
    eprint = {1901.10478},
    archivePrefix = {arXiv},
    primaryClass = {hep-ex}
}

@article{Castell_Mor_2020,
    author = {Castell\'o-Mor, N. and others},
    collaboration = {DAMIC-M},
    title = {{DAMIC-M experiment: Thick, silicon CCDs to search for light dark matter}},
    journal = {Nucl. Instrum. Meth. A},
    volume = {958},
    pages = {162933},
    year = {2020},
    doi = {10.1016/j.nima.2019.162933},
    eprint = {2001.01476},
    archivePrefix = {arXiv},
    primaryClass = {physics.ins-det}
}

@misc{aguilararevalo2022oscuraexperiment,
    author = {Aguilar-Arevalo, Alexis and others},
    collaboration = {Oscura},
    title = {{The Oscura Experiment}},
    journal = {arXiv e-prints},
    year = {2022},
    eprint = {2202.10518},
    archivePrefix = {arXiv},
    primaryClass = {astro-ph.IM}
}

@article{Nollett_2014,
    author = {Nollett, Kenneth M. and Steigman, Gary},
    title = {{BBN and the CMB constrain light, electromagnetically coupled WIMPs}},
    journal = {Phys. Rev. D},
    volume = {89},
    number = {8},
    pages = {083508},
    year = {2014},
    doi = {10.1103/PhysRevD.89.083508},
    eprint = {1305.1481},
    archivePrefix = {arXiv},
    primaryClass = {astro-ph.CO}
}

@article{B_hm_2013,
    author = {B{\oe}hm, C{\'e}line and others},
    title = {{A lower bound on the mass of cold thermal dark matter from Planck}},
    journal = {JCAP},
    volume = {08},
    pages = {041},
    year = {2013},
    doi = {10.1088/1475-7516/2013/08/041},
    eprint = {1303.6270},
    archivePrefix = {arXiv},
    primaryClass = {astro-ph.CO}
}

@article{Krnjaic_2020,
    author = {Krnjaic, Gordan and McDermott, Samuel D.},
    title = {{Implications of BBN bounds for cosmic ray upscattered dark matter}},
    journal = {Phys. Rev. D},
    volume = {101},
    number = {12},
    pages = {123022},
    year = {2020},
    doi = {10.1103/PhysRevD.101.123022},
    eprint = {1908.00007},
    archivePrefix = {arXiv},
    primaryClass = {hep-ph}
}

@article{Sabti_2020,
    author = {Sabti, Nashwan and others},
    title = {{Refined bounds on MeV-scale thermal dark sectors from BBN and the CMB}},
    journal = {JCAP},
    volume = {01},
    pages = {004},
    year = {2020},
    doi = {10.1088/1475-7516/2020/01/004},
    eprint = {1910.01649},
    archivePrefix = {arXiv},
    primaryClass = {hep-ph}
}

@article{Steigman_2012,
    author = {Steigman, Gary and Dasgupta, Basudeb and Beacom, John F.},
    title = {{Precise relic WIMP abundance and its impact on searches for dark matter annihilation}},
    journal = {Phys. Rev. D},
    volume = {86},
    number = {2},
    pages = {023506},
    year = {2012},
    doi = {10.1103/PhysRevD.86.023506},
    eprint = {1204.3622},
    archivePrefix = {arXiv},
    primaryClass = {hep-ph}
}

@article{Finkbeiner_2012,
    author = {Finkbeiner, Douglas P. and others},
    title = {{Searching for dark matter in the CMB: A compact parametrization of energy injection from new physics}},
    journal = {Phys. Rev. D},
    volume = {85},
    number = {4},
    pages = {043522},
    year = {2012},
    doi = {10.1103/PhysRevD.85.043522},
    eprint = {1109.6322},
    archivePrefix = {arXiv},
    primaryClass = {astro-ph.CO}
}

@article{2020,
    author = {Aghanim, N. and others},
    collaboration = {Planck},
    title = {{Planck 2018 results. VI. Cosmological parameters}},
    journal = {Astron. Astrophys.},
    volume = {641},
    pages = {A6},
    year = {2020},
    doi = {10.1051/0004-6361/201833910},
    eprint = {1807.06209},
    archivePrefix = {arXiv},
    primaryClass = {astro-ph.CO}
}

@article{GONDOLO1991145,
    author = {Gondolo, Paolo and Gelmini, Graciela},
    title = {{Cosmic abundances of stable particles: Improved analysis}},
    journal = {Nucl. Phys. B},
    volume = {360},
    number = {1},
    pages = {145--179},
    year = {1991},
    doi = {10.1016/0550-3213(91)90438-4}
}

@article{Li_2023,
    author = {Li, Shuaijie and others},
    collaboration = {PandaX},
    title = {{Search for Light Dark Matter with Ionization Signals in the PandaX-4T Experiment}},
    doi = {10.1103/physrevlett.130.261001},
    journal = {Phys. Rev. Lett.},
    volume = {130},
    number = {26},
    pages = {261001},
    year = {2023},
    eprint = {2212.10067},
    archivePrefix = {arXiv},
    primaryClass = {hep-ex}
}

@article{Agnes_2023,
    author = {Agnes, P. and others},
    collaboration = {DarkSide-50},
    title = {{Search for Dark Matter Particle Interactions with Electron Final States with DarkSide-50}},
    journal = {Phys. Rev. Lett.},
    volume = {130},
    number = {10},
    pages = {101002},
    year = {2023},
    doi = {10.1103/physrevlett.130.101002},
    eprint = {2207.11968},
    archivePrefix = {arXiv},
    primaryClass = {hep-ex}
}

@article{An_2021,
    author = {An, Haipeng and others},
    title = {{Solar reflection of dark matter}},
    journal = {Phys. Rev. D},
    volume = {104},
    number = {10},
    pages = {103026},
    year = {2021},
    doi = {10.1103/physrevd.104.103026},
    eprint = {2108.10332},
    archivePrefix = {arXiv},
    primaryClass = {hep-ph}
}

@article{Abdelhameed_2019,
    author = {Abdelhameed, A. H. and others},
    collaboration = {CRESST},
    title = {{First results from the CRESST-III low-mass dark matter program}},
    journal = {Phys. Rev. D},
    volume = {100},
    number = {10},
    pages = {102002},
    year = {2019},
    doi = {10.1103/physrevd.100.102002},
    eprint = {1904.00498},
    archivePrefix = {arXiv},
    primaryClass = {astro-ph.CO}
}

@misc{cresstcollaboration2024observationsinglephotonscresst,
    author = {Angloher, Godehard and others},
    collaboration = {CRESST},
    title={First observation of single photons in a CRESST detector and new dark matter exclusion limits}, 
    year = {2024},
    eprint = {2405.06527},
    archivePrefix = {arXiv},
    primaryClass = {astro-ph.CO}
}

@article{Essig_2013,
    author = {Essig, Rouven and others},
    title = {{Constraining light dark matter with low-energy e+e- colliders}},
    journal = {JHEP},
    volume = {11},
    pages = {167},
    year = {2013},
    eprint = {1309.5084},
    archivePrefix = {arXiv},
    primaryClass = {hep-ph},
    doi = {10.1007/jhep11(2013)167}
}

@article{Izaguirre_2013,
    author = {Izaguirre, Eder and others},
    title = {{New electron beam-dump experiments to search for MeV to few-GeV dark matter}},
    journal = {Phys. Rev. D},
    volume = {88},
    number = {11},
    pages = {114015},
    year = {2013},
    doi = {10.1103/physrevd.88.114015}
}

@article{CCFR:1991lpl,
    author = {Mishra, S. R. and others},
    collaboration = {CCFR},
    title = {{Neutrino tridents and W Z interference patterns in elastic nu e scattering}},
    journal = {Phys. Rev. Lett.},
    volume = {66},
    pages = {3117--3120},
    year = {1991},
    doi = {10.1103/PhysRevLett.66.3117},
    reportNumber = {NEVIS-1459, FERMILAB-PUB-90-256-E}
}

@article{Altmannshofer:2014pba,
    author = {Altmannshofer, Wolfgang and others},
    title = {{Neutrino Trident Production: A Powerful Probe of Missing Leptonics}},
    journal = {Phys. Rev. Lett.},
    volume = {113},
    number = {9},
    pages = {091801},
    year = {2014},
    doi = {10.1103/PhysRevLett.113.091801},
    eprint = {1406.2332},
    archivePrefix = {arXiv},
    primaryClass = {hep-ph}
}

@misc{krnjaic2025testingthermalrelicdarkmatter,
    author = {Krnjaic, Gordan},
    title = {{Testing Thermal-Relic Dark Matter with a Dark Photon Mediator}},
    year = {2025},
    eprint = {2505.04626},
    archivePrefix = {arXiv},
    primaryClass = {hep-ph}
}

@article{DarkSide:2022dhx,
    author = "Agnes, P. and others",
    collaboration = "DarkSide",
    title = "{Search for Dark-Matter{\textendash}Nucleon Interactions via Migdal Effect with DarkSide-50}",
    eprint = "2207.11967",
    archivePrefix = "arXiv",
    primaryClass = "hep-ex",
    doi = "10.1103/PhysRevLett.130.101001",
    journal = "Phys. Rev. Lett.",
    volume = "130",
    number = "10",
    pages = "101001",
    year = "2023"
}

@article{COSINE-100:2021poy,
    author = "Adhikari, G. and others",
    collaboration = "COSINE-100",
    title = "{Searching for low-mass dark matter via the Migdal effect in COSINE-100}",
    eprint = "2110.05806",
    archivePrefix = "arXiv",
    primaryClass = "hep-ex",
    doi = "10.1103/PhysRevD.105.042006",
    journal = "Phys. Rev. D",
    volume = "105",
    number = "4",
    pages = "042006",
    year = "2022"
}

@article{Baxter:2019pnz,
    author = "Baxter, Daniel and Kahn, Yonatan and Krnjaic, Gordan",
    title = "{Electron Ionization via Dark Matter-Electron Scattering and the Migdal Effect}",
    eprint = "1908.00012",
    archivePrefix = "arXiv",
    primaryClass = "hep-ph",
    reportNumber = "FERMILAB-PUB-19-257-A",
    doi = "10.1103/PhysRevD.101.076014",
    journal = "Phys. Rev. D",
    volume = "101",
    number = "7",
    pages = "076014",
    year = "2020"
}

@misc{Xu:2026acq,
    author        = "Xu, Jingke and Kim, Jeonghwa and Adams, Duncan and Lenardo, Brian G. and Lippincott, Walter Hugh and Essig, Rouven",
    title         = "{Nuclear Recoil Migdal Effect in Liquid Xenon Dark Matter Experiments}",
    eprint        = "2606.03174",
    archivePrefix = "arXiv",
    primaryClass  = "hep-ex",
    reportNumber  = "LLNL-JRNL-2017344",
    month         = jun,
    year          = "2026"
}

@article{Yi:2026fmf,
    author = "Yi, Difan and others",
    title = "{Direct observation of the Migdal effect induced by neutron bombardment}",
    doi = "10.1038/s41586-025-09918-8",
    journal = "Nature",
    volume = "649",
    number = "8097",
    pages = "580--583",
    year = "2026"
}

@article{Adams:2022zvg,
    author = "Adams, Duncan and Baxter, Daniel and Day, Hannah and Essig, Rouven and Kahn, Yonatan",
    title = "{Measuring the Migdal effect in semiconductors for dark matter detection}",
    eprint = "2210.04917",
    archivePrefix = "arXiv",
    primaryClass = "hep-ph",
    reportNumber = "FERMILAB-PUB-22-705-PPD-QIS-T",
    doi = "10.1103/PhysRevD.107.L041303",
    journal = "Phys. Rev. D",
    volume = "107",
    number = "4",
    pages = "L041303",
    year = "2023"
}

@article{MIGDAL:2022yip,
    author = "Ara{\'u}jo, H. M. and others",
    collaboration = "MIGDAL",
    title = "{The MIGDAL experiment: Measuring a rare atomic process to aid the search for dark matter}",
    eprint = "2207.08284",
    archivePrefix = "arXiv",
    primaryClass = "hep-ex",
    doi = "10.1016/j.astropartphys.2023.102853",
    journal = "Astropart. Phys.",
    volume = "151",
    pages = "102853",
    year = "2023"
}

@article{Xu:2023wev,
    author = "Xu, Jingke and others",
    title = "{Search for the Migdal effect in liquid xenon with keV-level nuclear recoils}",
    eprint = "2307.12952",
    archivePrefix = "arXiv",
    primaryClass = "hep-ex",
    reportNumber = "LLNL-JRNL-850170",
    doi = "10.1103/PhysRevD.109.L051101",
    journal = "Phys. Rev. D",
    volume = "109",
    number = "5",
    pages = "L051101",
    year = "2024"
}

@article{Armengaud_2022,
    author = "Armengaud, E. and others",
    collaboration = "EDELWEISS",
    title = "{Search for sub-GeV dark matter via the Migdal effect with an EDELWEISS germanium detector with NbSi transition-edge sensors}",
    eprint = "2203.03993",
    archivePrefix = "arXiv",
    primaryClass = "hep-ex",
    doi = "10.1103/PhysRevD.106.062004",
    journal = "Phys. Rev. D",
    volume = "106",
    number = "6",
    pages = "062004",
    year = "2022"
}

@article{CDEX:2019hzn,
    author = "Liu, Z. Z. and others",
    collaboration = "CDEX",
    title = "{Constraints on Spin-Independent Nucleus Scattering with sub-GeV Weakly Interacting Massive Particle Dark Matter from the CDEX-1B Experiment at the China Jinping Underground Laboratory}",
    eprint = "1905.00354",
    archivePrefix = "arXiv",
    primaryClass = "hep-ex",
    doi = "10.1103/PhysRevLett.123.161301",
    journal = "Phys. Rev. Lett.",
    volume = "123",
    number = "16",
    pages = "161301",
    year = "2019"
}

@article{XENON:2019zpr,
    author = "Aprile, E. and others",
    collaboration = "XENON",
    title = "{Search for Light Dark Matter Interactions Enhanced by the Migdal Effect or Bremsstrahlung in XENON1T}",
    eprint = "1907.12771",
    archivePrefix = "arXiv",
    primaryClass = "hep-ex",
    doi = "10.1103/PhysRevLett.123.241803",
    journal = "Phys. Rev. Lett.",
    volume = "123",
    number = "24",
    pages = "241803",
    year = "2019"
}

@article{PandaX:2023xgl,
    author = "Huang, Di and others",
    collaboration = "PandaX",
    title = "{Search for Dark-Matter{\textendash}Nucleon Interactions with a Dark Mediator in PandaX-4T}",
    eprint = "2308.01540",
    archivePrefix = "arXiv",
    primaryClass = "hep-ex",
    doi = "10.1103/PhysRevLett.131.191002",
    journal = "Phys. Rev. Lett.",
    volume = "131",
    number = "19",
    pages = "191002",
    year = "2023"
}

@article{Dolan:2017xbu,
    author = "Dolan, Matthew J. and Kahlhoefer, Felix and McCabe, Christopher",
    title = "{Directly detecting sub-GeV dark matter with electrons from nuclear scattering}",
    eprint = "1711.09906",
    archivePrefix = "arXiv",
    primaryClass = "hep-ph",
    reportNumber = "KCL-PH-TH/2017-54, TTK-17-43, KCL-PH-TH-2017-54",
    doi = "10.1103/PhysRevLett.121.101801",
    journal = "Phys. Rev. Lett.",
    volume = "121",
    number = "10",
    pages = "101801",
    year = "2018"
}

@article{Ibe:2017yqa,
    author = "Ibe, Masahiro and Nakano, Wakutaka and Shoji, Yutaro and Suzuki, Kazumine",
    title = "{Migdal Effect in Dark Matter Direct Detection Experiments}",
    eprint = "1707.07258",
    archivePrefix = "arXiv",
    primaryClass = "hep-ph",
    reportNumber = "IPMU17-0100",
    doi = "10.1007/JHEP03(2018)194",
    journal = "JHEP",
    volume = "03",
    pages = "194",
    year = "2018"
}

@article{XENON:2017vdw,
    author = "Aprile, E. and others",
    collaboration = "XENON",
    title = "{First Dark Matter Search Results from the XENON1T Experiment}",
    eprint = "1705.06655",
    archivePrefix = "arXiv",
    primaryClass = "astro-ph.CO",
    doi = "10.1103/PhysRevLett.119.181301",
    journal = "Phys. Rev. Lett.",
    volume = "119",
    number = "18",
    pages = "181301",
    year = "2017"
}

@article{Altmannshofer:2019xda,
    author = "Altmannshofer, Wolfgang and Davighi, Joe and Nardecchia, Marco",
    title = "{Gauging the accidental symmetries of the standard model, and implications for the flavor anomalies}",
    eprint = "1909.02021",
    archivePrefix = "arXiv",
    primaryClass = "hep-ph",
    doi = "10.1103/PhysRevD.101.015004",
    journal = "Phys. Rev. D",
    volume = "101",
    number = "1",
    pages = "015004",
    year = "2020"
}

@article{Dror:2017nsg,
    author = "Dror, Jeff A. and Lasenby, Robert and Pospelov, Maxim",
    title = "{Dark forces coupled to nonconserved currents}",
    eprint = "1707.01503",
    archivePrefix = "arXiv",
    primaryClass = "hep-ph",
    doi = "10.1103/PhysRevD.96.075036",
    journal = "Phys. Rev. D",
    volume = "96",
    number = "7",
    pages = "075036",
    year = "2017"
}

@article{Dror:2017ehi,
    author = "Dror, Jeff A. and Lasenby, Robert and Pospelov, Maxim",
    title = "{New constraints on light vectors coupled to anomalous currents}",
    eprint = "1705.06726",
    archivePrefix = "arXiv",
    primaryClass = "hep-ph",
    doi = "10.1103/PhysRevLett.119.141803",
    journal = "Phys. Rev. Lett.",
    volume = "119",
    number = "14",
    pages = "141803",
    year = "2017"
}

@article{Dobrescu:2014fca,
    author = "Dobrescu, Bogdan A. and Frugiuele, Claudia",
    title = "{Hidden GeV-Scale Interactions of Quarks}",
    eprint = "1404.3947",
    archivePrefix = "arXiv",
    primaryClass = "hep-ph",
    reportNumber = "FERMILAB-PUB-14-098-T",
    doi = "10.1103/PhysRevLett.113.061801",
    journal = "Phys. Rev. Lett.",
    volume = "113",
    pages = "061801",
    year = "2014"
}

@article{Kahn:2016vjr,
    author = "Kahn, Yonatan and Krnjaic, Gordan and Mishra-Sharma, Siddharth and Tait, Tim M. P.",
    title = "{Light Weakly Coupled Axial Forces: Models, Constraints, and Projections}",
    eprint = "1609.09072",
    archivePrefix = "arXiv",
    primaryClass = "hep-ph",
    reportNumber = "FERMILAB-PUB-16-385-PPD, UCI-HEP-TR-2016-15, MITP-16-098, PUPT-2507",
    doi = "10.1007/JHEP05(2017)002",
    journal = "JHEP",
    volume = "05",
    pages = "002",
    year = "2017"
}

@book{Fabbrichesi_2021,
    author = {Fabbrichesi, Marco and Gabrielli, Emidio and Lanfranchi, Gaia},
    publisher = {Springer International Publishing},
    title = {{The Physics of the Dark Photon: A Primer}},
    series = {SpringerBriefs in Physics},
    year = {2021},
    doi = {10.1007/978-3-030-62519-1}
}

@article{PhysRevD.111.056003,
    title = {Radiative corrections to light thermal pseudo-Dirac dark matter},
    author = {Mohlabeng, Gopolang and Mondol, Adreja and Tait, Tim M. P.},
    journal = {Phys. Rev. D},
    volume = {111},
    issue = {5},
    pages = {056003},
    numpages = {10},
    year = {2025},
    eprint = {2405.08881},
    archivePrefix = {arXiv},
    primaryClass = {hep-ph},
    month = {Mar},
    publisher = {American Physical Society},
    doi = {10.1103/PhysRevD.111.056003},
    url = {https://link.aps.org/doi/10.1103/PhysRevD.111.056003}
}

@article{KA:2023dyz,
    author = "K. A., ShivaSankar and Das, Arindam and Lambiase, Gaetano and Nomura, Takaaki and Orikasa, Yuta",
    title = "{Probing chiral and flavored $Z^\prime $ from cosmic bursts through neutrino interactions}",
    eprint = "2308.14483",
    archivePrefix = "arXiv",
    primaryClass = "hep-ph",
    doi = "10.1140/epjc/s10052-024-13530-x",
    journal = "Eur. Phys. J. C",
    volume = "84",
    number = "11",
    pages = "1224",
    year = "2024"
}

@article{Barman:2024lxy,
    author = "Barman, Basabendu and Das, Arindam and Mandal, Sanjoy",
    title = "{Dark matter-electron scattering and freeze-in scenarios in the light of Z' mediation}",
    eprint = "2407.00969",
    archivePrefix = "arXiv",
    primaryClass = "hep-ph",
    doi = "10.1103/PhysRevD.110.055029",
    journal = "Phys. Rev. D",
    volume = "110",
    number = "5",
    pages = "055029",
    year = "2024"
}

@misc{Barman:2025jbo,
    author = "Barman, Basabendu and Bhattacharjee, Pooja and Das, Arindam",
    title = "{New constraints on $Z^\prime$ from captured dark matter annihilation in astrophysical objects}",
    eprint = "2509.12192",
    archivePrefix = "arXiv",
    primaryClass = "hep-ph",
    month = "9",
    year = "2025"
}

@article{Bernal:2025szh,
    author = "Bernal, Nicol{\'a}s and Neto, Jacinto P. and Silva-Malpartida, Javier and Queiroz, Farinaldo S.",
    title = "{Enabling thermal dark matter within the vanilla L{\ensuremath{\mu}}-L{\ensuremath{\tau}} model}",
    eprint = "2507.02048",
    archivePrefix = "arXiv",
    primaryClass = "hep-ph",
    doi = "10.1103/8g85-c8sh",
    journal = "Phys. Rev. D",
    volume = "112",
    number = "7",
    pages = "075042",
    year = "2025"
}

@article{Carew:2023qrj,
    author = "Carew, Ben and Caddell, Ashlee R. and Maity, Tarak Nath and O'Hare, Ciaran A. J.",
    title = "{Neutrino fog for dark matter-electron scattering experiments}",
    eprint = "2312.04303",
    archivePrefix = "arXiv",
    primaryClass = "hep-ph",
    doi = "10.1103/PhysRevD.109.083016",
    journal = "Phys. Rev. D",
    volume = "109",
    number = "8",
    pages = "083016",
    year = "2024"
}

@article{Boehm:2003hm,
    author = "Boehm, C. and Fayet, P.",
    title = "{Scalar dark matter candidates}",
    eprint = "hep-ph/0305261",
    archivePrefix = "arXiv",
    doi = "10.1016/j.nuclphysb.2004.01.015",
    journal = "Nucl. Phys. B",
    volume = "683",
    pages = "219--263",
    year = "2004"
}

@article{Essig:2011nj,
    author = "Essig, Rouven and Mardon, Jeremy and Volansky, Tomer",
    title = "{Direct Detection of Sub-GeV Dark Matter}",
    eprint = "1108.5383",
    archivePrefix = "arXiv",
    primaryClass = "hep-ph",
    doi = "10.1103/PhysRevD.85.076007",
    journal = "Phys. Rev. D",
    volume = "85",
    pages = "076007",
    year = "2012"
}

@article{Essig:2012yx,
    author = "Essig, Rouven and Manalaysay, Aaron and Mardon, Jeremy and Sorensen, Peter and Volansky, Tomer",
    title = "{First Direct Detection Limits on sub-GeV Dark Matter from XENON10}",
    eprint = "1206.2644",
    archivePrefix = "arXiv",
    primaryClass = "astro-ph.CO",
    doi = "10.1103/PhysRevLett.109.021301",
    journal = "Phys. Rev. Lett.",
    volume = "109",
    pages = "021301",
    year = "2012"
}

@article{Essig:2015cda,
    author = "Essig, Rouven and Fernandez-Serra, Marivi and Mardon, Jeremy and Soto, Adrian and Volansky, Tomer and Yu, Tien-Tien",
    title = "{Direct Detection of sub-GeV Dark Matter with Semiconductor Targets}",
    eprint = "1509.01598",
    archivePrefix = "arXiv",
    primaryClass = "hep-ph",
    doi = "10.1007/JHEP05(2016)046",
    journal = "JHEP",
    volume = "05",
    pages = "046",
    year = "2016"
}

@article{Chen:2003gz,
    author = "Chen, Xuelei and Kamionkowski, Marc",
    title = "{Particle decays during the cosmic dark ages}",
    eprint = "astro-ph/0310473",
    archivePrefix = "arXiv",
    doi = "10.1103/PhysRevD.70.043502",
    journal = "Phys. Rev. D",
    volume = "70",
    pages = "043502",
    year = "2004"
}

@article{Padmanabhan:2005es,
    author = "Padmanabhan, Nikhil and Finkbeiner, Douglas P.",
    title = "{Detecting dark matter annihilation with CMB polarization: Signatures and experimental prospects}",
    eprint = "astro-ph/0503486",
    archivePrefix = "arXiv",
    doi = "10.1103/PhysRevD.72.023508",
    journal = "Phys. Rev. D",
    volume = "72",
    pages = "023508",
    year = "2005"
}

@article{Galli:2009zc,
    author = "Galli, S. and Iocco, F. and Bertone, G. and Melchiorri, A.",
    title = "{CMB constraints on Dark Matter models with large annihilation cross-section}",
    eprint = "0905.0003",
    archivePrefix = "arXiv",
    primaryClass = "astro-ph.CO",
    doi = "10.1103/PhysRevD.80.023505",
    journal = "Phys. Rev. D",
    volume = "80",
    pages = "023505",
    year = "2009"
}

@article{SuperCDMS:2018mne,
    author = "Agnese, R. and others",
    collaboration = "SuperCDMS",
    title = "{First Dark Matter Constraints from a SuperCDMS Single-Charge Sensitive Detector}",
    eprint = "1804.10697",
    archivePrefix = "arXiv",
    primaryClass = "hep-ex",
    doi = "10.1103/PhysRevLett.121.051301",
    journal = "Phys. Rev. Lett.",
    volume = "121",
    number = "5",
    pages = "051301",
    year = "2018"
}

@article{EDELWEISS:2019vjv,
    author = "Armengaud, E. and others",
    collaboration = "EDELWEISS",
    title = "{Searching for low-mass dark matter particles with a massive Ge bolometer operated above-ground}",
    eprint = "1901.03588",
    archivePrefix = "arXiv",
    primaryClass = "astro-ph.GA",
    doi = "10.1103/PhysRevD.99.082003",
    journal = "Phys. Rev. D",
    volume = "99",
    number = "8",
    pages = "082003",
    year = "2019"
}

@article{SENSEI:2023gie,
    author = "Adari, Prakruth and others",
    collaboration = "SENSEI",
    title = "{First Direct-Detection Results on Sub-GeV Dark Matter Using the SENSEI Detector at SNOLAB}",
    eprint = "2312.13342",
    archivePrefix = "arXiv",
    primaryClass = "astro-ph.CO",
    doi = "10.1103/PhysRevLett.134.011804",
    journal = "Phys. Rev. Lett.",
    volume = "134",
    number = "1",
    pages = "011804",
    year = "2025"
}

@article{XENON:2024xur,
    author = "Aprile, E. and others",
    collaboration = "XENON",
    title = "{Search for Light Dark Matter in Low-Energy Ionization Signals from XENONnT}",
    eprint = "2411.15289",
    archivePrefix = "arXiv",
    primaryClass = "hep-ex",
    doi = "10.1103/PhysRevLett.134.161004",
    journal = "Phys. Rev. Lett.",
    volume = "134",
    number = "16",
    pages = "161004",
    year = "2025"
}

@article{Chu:2023fmc,
    author = "Chu, Xiaoyong and Pradler, Josef",
    title = "{On the minimal mass of thermal dark matter and the viability of millicharged particles affecting 21-cm cosmology}",
    eprint = "2310.06611",
    archivePrefix = "arXiv",
    primaryClass = "hep-ph",
    doi = "10.1103/PhysRevD.109.103510",
    journal = "Phys. Rev. D",
    volume = "109",
    number = "10",
    pages = "103510",
    year = "2024"
}

@article{Chu:2022qnb,
    author = "Chu, Xiaoyong and Kuo, Jui-Lin and Pradler, Josef",
    title = "{Towards a full description of MeV dark matter decoupling: A self-consistent determination of relic abundance and $N_{\rm eff}$}",
    eprint = "2205.05714",
    archivePrefix = "arXiv",
    primaryClass = "hep-ph",
    doi = "10.1103/PhysRevD.106.055022",
    journal = "Phys. Rev. D",
    volume = "106",
    number = "5",
    pages = "055022",
    year = "2022"
}

\end{document}